\documentclass{appolb}
\usepackage{epsf}
\begin{document}
\setcounter{page}{3647}
\title{Skewed   Parton Distributions%
\thanks{Presented at XXXIX Cracow School on Theoretical Physics}%
}
\author{A.V. RADYUSHKIN\footnote{Also at Laboratory of Theoretical Physics,
JINR, Dubna, Russia
}
\address{Physics Department, Old Dominion University, \\
Norfolk, VA 23529, USA \\ and \\ 
Jefferson Lab,
  Newport News,VA 23606, USA}}
\maketitle
\begin{abstract}
Applications of perturbative QCD to
deeply virtual Compton scattering
and hard exclusive electroproduction 
processes require a generalization of 
usual parton distributions for the case when
long-distance information
is accumulated in  nonforward matrix
elements 
of quark and gluon light-cone  operators.
We describe 
two types of nonperturbative functions 
parametrizing such matrix elements:
double distributions $F(x,y;t)$
and skewed distribution functions
${\cal F}_{\zeta}(X;t)$,   
discuss their  properties,
and 
basic uses in the QCD description of 
hard exclusive processes.

\end{abstract}

\section{Introduction}

 The standard feature of  
applications of perturbative QCD to hard
processes  is the introduction of  phenomenological
functions  accumulating information about
nonperturbative long-distance dynamics.    The well-known 
 examples are the 
 parton distribution functions $f_{p/H}(x)$ \cite{feynman}
used in  perturbative QCD approaches to hard
inclusive processes and 
distribution amplitudes   $\varphi_{\pi}(x), \,
\varphi_{N}(x_1 , x_2 , x_3) $,  which naturally
emerge in the  asymptotic  QCD
analyses of hard exclusive processes 
\cite{cz77,farjack,ar77,tmf,bl,bl80}.

The cases of 
deeply virtual Compton scattering (DVCS) 
and hard exclusive electroproduction 
processes \cite{ji,compton,gluon,npd,cfs,drm} 
involve nonforward matrix
elements $\langle p' | \ldots | p \rangle$. 
The important feature (noticed  long ago  \cite{barloe,glr})
 is that  kinematics of hard elastic 
electroproduction processes (DVCS can be also treated as one of them) 
requires 
the presence of the longitudinal component
in the  momentum transfer 
$r\equiv p - p'$ from the initial hadron to the final:
$r^+ = \zeta p^+$. For DVCS and $\rho$-electroproduction 
in the region       $Q^2 \gg  |t|, m_H^2$,
the longitudinal momentum asymmetry (or ``skewedness'') 
parameter $\zeta$
coincides with  the
Bjorken variable $x_{Bj} = Q^2/2(pq)$ associated with 
the virtual photon momentum $q$ \cite{afs}. 
This means that the nonperturbative matrix element
$\langle p' | \ldots | p \rangle$ 
is nonsymmetric (skewed), 
 and the distributions which appear in the 
hard elastic electroproduction amplitudes
differ  from those studied in inclusive 
processes. In the latter case, 
one has  a symmetric situation when the same
momentum $p$ appears in both brackets of the hadronic matrix element
$\langle p | \ldots | p \rangle$.

To parametrize  nonforward matrix elements $\langle p-r \,  |     \,   
{\cal O}(0,z)  \,  |     \, p \rangle \,  |     \, _{z^2=0}$ 
of quark and gluon light-cone  operators
     one can use 
two basic  types of nonperturbative functions. 
The   double distributions (DDs) $F(x,y;t)$ \cite{compton,npd,ddee,sssdd}   
specify the Sudakov
light-cone ``plus'' fractions $xp^+$ and $yr^+$ 
of the initial hadron momentum $p$ and the momentum transfer $r$ 
carried by the initial parton.   
The other possibility is to treat
 the proportionality  coefficient 
$\zeta$ as an 
independent parameter  and    introduce 
an alternative description in terms
of the   nonforward parton distributions (NFPDs) 
${\cal F}_{\zeta}(X;t)$ 
with $X=x+y \zeta$ being the total 
fraction of the initial hadron momentum 
taken  by the initial  parton.
The shape of  NFPDs  explicitly
depends on the parameter $\zeta$ characterizing the {\it skewedness}
of the relevant nonforward matrix element.
This parametrization of  nonforward matrix 
elements  by ${\cal F}_{\zeta}(X;t)$
is similar to that proposed  by 
X. Ji \cite{ji} who introduced 
 off-forward parton distributions (OFPDs) $H(\tilde x,\xi;t)$
in which the parton momenta and  the skewedness
parameter $\xi\equiv r^+ / 2 P^+$ 
are measured in units of the average 
hadron momentum $P=(p+p')/2$. 
  OFPDs and NFPDs \cite{npd,cfs}
can be   treated 
as particular forms  of {\it skewed } parton 
distributions (SPDs).
One can also introduce the version of  DDs (``$\alpha$-DDs'' \cite{sssdd}) 
in which the active parton momentum is  written in terms of symmetric 
variables  
$k= xP + (1+\alpha) r/2$.

The basics of the    PQCD approaches   
incorporating  skewed   parton distributions 
were formulated in refs.\cite{ji,compton,gluon,npd}.
A detailed analysis of PQCD factorization 
for hard meson electroproduction processes
was given in Ref.  \cite{cfs}.
Our goal in the present lectures  is to give a 
description of the approach outlined in 
our  papers \cite{compton,gluon,npd,ddee,sssdd}.

\section{Double distributions and their symmetries}

In the pQCD factorization treatment of hard
electroproduction processes, the 
nonperturbative information is accumulated in the
nonforward matrix elements
$\langle p-r \,  |     \,  {\cal O} (0,z)  \,  |     \,  p \rangle $
of   light cone operators $ {\cal O} (0,z) $. 
For  $z^2=0$ the matrix elements 
depend on the relative coordinate $z$
through two Lorentz invariant variables $(pz)$ and $(rz)$.
In the forward case, when $r=0$, 
one obtains the usual quark helicity-averaged densities   
by  Fourier transforming   the relevant  matrix element 
with respect to $(pz)$ 
\begin{eqnarray} 
&& \langle p,s'\, \,  |     \,  \, \bar \psi_a(0) \hat z \label{33}
E(0,z;A)  \psi_a(z) \, \,  |     \,  \, p,s \rangle \,  |     \, _{z^2=0} 
\\ && =  \bar u(p,s')  \hat z u(p,s)  
   \int_0^1  \,  
 \left ( e^{-ix(pz)}f_a(x) 
  -   e^{ix(pz)}f_{\bar a}(x)
\right ) \, dx \, , 
\nonumber   \end{eqnarray}
where $E(0,z;A)$ is the gauge link,  
 $\bar u(p',s'), u(p,s)$ are the Dirac
spinors and we use the  notation
$\gamma_{\alpha} z^{\alpha} \equiv \hat z$.

\begin{figure}[htb]
\mbox{
   \epsfxsize=13.5cm
 \epsfysize=5cm
  \epsffile{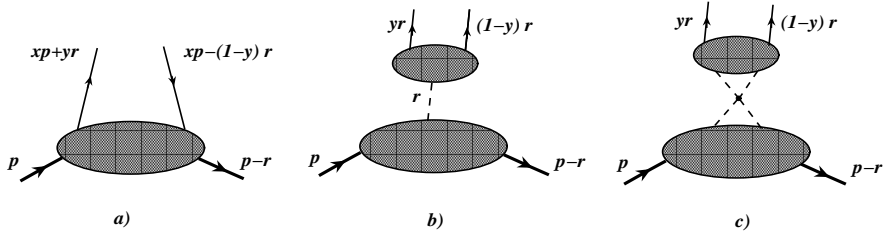}  
   }
  \vspace{-2cm}
{\caption{\label{fg:exchange} $a)$ Parton picture
in terms of $y$-DDs; $b,c)$   $F_M$-type contributions. 
   }}
\end{figure}

The parameter $x$ in this representation has an evident 
interpretation: it characterizes which
the fraction of the initial hadron  momentum is carried 
by the active parton.

In the nonforward case,  we can  
use the  double Fourier  representation 
with respect to both $(pz)$ and $(rz)$: 
\begin{eqnarray} 
&& \langle p',s'\, \,  |     \,  \, \bar \psi_a(0) \hat z 
E(0,z;A)  \psi_a(z) \, \,  |     \,  \, p,s \rangle \,  | 
    \, _{z^2=0} 
\label{31}  \\ &&  =   
\int_0^1  dy   \int_{-1}^1  \,  
  e^{-ix(pz)-iy(r z)} \,\left. \biggl \{ 
   \bar u(p',s')  \hat z u(p,s) \, \tilde F_a(x,y;t) \right. 
 \nonumber   \\ && \left.  \, +
\, \bar u(p',s')  \frac {\hat z  \hat r - \hat r \hat z}{4M}
 u(p,s) \, \tilde K_a(x,y;t) \, \right. \biggr \} \, 
 \theta( 0 \leq x+y \leq 1) \, dx   \nonumber    \\
 && + \frac{(zr) }{m}\,  \bar u(p',s') u(p,s) \, 
 \int_0^1 e^{-iy(r z)}\, \Psi_a (y;t) \,  dy \, , 
\nonumber 
 \end{eqnarray} 
where $M$ is the nucleon mass and $s,s'$ specify the nucleon
 polarization. We use the  ``hat''
(rather than ``slash'') convention $\hat z \equiv z^{\mu} 
\gamma_{\mu}$. 
The parametrization of  nonforward matrix elements
must  include both the nonflip
term described here by  the functions 
$ F_a(x,y;t)$
and the spin-flip term 
 characterized by the functions  $ K_a(x,y;t)$. 
 
 \newpage
 
\begin{figure}[htb]
\mbox{
   \epsfxsize=13.5cm
 \epsfysize=7.5cm
 %\hspace{1.5cm}  
  \epsffile{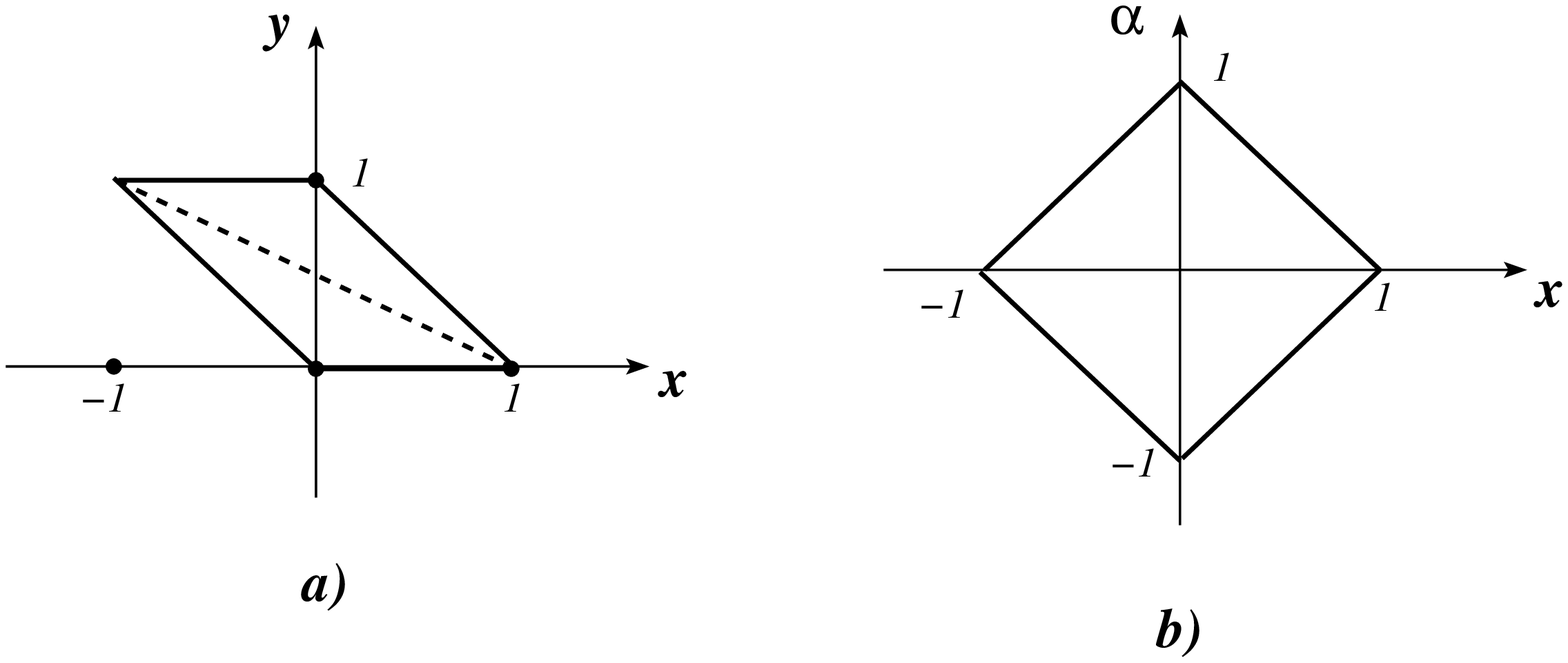}  }
  \vspace{-2.5cm} 
{\caption{\label{fg:support} $a)$ Support region 
and symmetry line  $y = \bar x/2$ for 
$y$-DDs 
$\tilde F(x,y;t)$; $b)$ support region 
for $\alpha$-DDs $\tilde f (x, \alpha)$. 
   }}
\end{figure}

 The parameters $x,y$ tell us that the active parton
 carries  the fractions $x$ of 
 the initial momentum $p$ and the fraction $y$ of 
 the momentum transfer $r$. 
Using the approach  \cite{spectral} based on the 
$\alpha$-representation analysis 
it is possible to  prove \cite{npd} that  double distributions
$F(x,y)$ have a natural property that both $x$ and $y$ 
satisfy the ``parton'' constraints 
 $0 \leq x \leq 1$, $ 0 \leq y \leq 1$
for  any Feynman  diagram contributing  to $F(x,y)$. 
A less obvious restriction   $0 \leq x+y \leq 1$  guarantees 
that the argument $X=x+y \zeta$ of the  skewed  distribution
${\cal F}_{\zeta} (X)$ also changes between the limits 
 The support area 
for  the {\it double distribution} $ \tilde F_a(x,y;t)$ 
is shown on Fig.\ref{fg:support}a.

\vspace{-2cm}

 In principle, we cannot exclude also 
the   possibility that  the functions 
$\tilde F (x,y;t)$    have  singular terms at $x=0$ 
proportional 
to $\delta (x)$ or its derivative(s).
Such terms  
have no projection  onto the usual parton densities.
We will denote them by $F_M (x,y;t)$ $-$  they may 
be interpreted as coming from the $t$-channel 
meson-exchange type contributions (see Fig.\ref{fg:exchange}b).  
In this case, the partons  just share 
the plus component of the momentum transfer $r$:
information about 
the magnitude of the  initial hadron momentum
is lost if the exchanged particle can be described
by a pole propagator $\sim 1/(t-m_M^2)$.
Hence, the meson-exchange  contributions to a double distribution
may look like
\begin{equation}
\tilde F_M^+(x,y;t) \sim  \delta (x) \,
 \frac{\varphi_{M}^+ (y)}{m_M^2 -t } \ \  \ 
{\rm or }  \ \  \  \tilde  F_M^-(x,y;t) 
\sim  \delta ' (x) \, \frac{\varphi^-_{M
}(y)}{m_M^2 -t } \ \  , \  \  {\rm etc.} \, , 
\end{equation} 
where $\varphi_M^{\pm}  (y)$ are  the functions related to the 
distribution amplitudes of the relevant mesons $M^{\pm}$.
The  two examples above correspond to $x$-even and $x$-odd
parts of the double distribution $\tilde F(x,y ;t)$. 
Another type of    terms in 
which the dependence on $(pz)$ is  lost can  be produced by 
diagrams containing  
a quartic pion vertex (Fig.\ref{fg:exchange}c).
 As shown by Polyakov and Weiss \cite{poweiss},
such terms  
correspond to an 
independent   $(rz) \bar u(p',s')    u(p,s) \Phi ((rz))$ 
type contribution which can be parametrized by a single integral
over $y$ involving an  
effective distribution amplitude  $\Psi (y;t)$.
The meson-exchange terms in $F(x,y ;t)$ and $K(x,y ;t)$ 
as well as 
Polyakov-Weiss terms are invisible 
in the forward limit, hence the existing 
knowledge of the usual parton densities 
cannot be used  to   constrain
these terms. Later,  describing 
the models for skewed distributions, 
we discuss  only the ``forward visible parts''
of SPDs which are obtained by scanning 
the $x \neq 0$ parts of the relevant DDs.

Comparing Eq. (\ref{33}) 
with the $r =0$ limit of the DD definition (\ref{31})
gives  the   
``reduction formulas'' 
relating   the double distribution $\tilde F_{a}(x,y;t=0)$  
to the quark and antiquark parton  densities
\begin{equation} 
\int_0^{1-x} \tilde F_{a}(x,y;t=0)|_{x>0} \, dy= 
 f_{a}(x) \ ; \ 
\int_{-x}^1  \tilde F_{ a}(x,y;t=0)|_{x< 0}\, dy= 
-  f_{\bar a}(-x)   \label{eq:redfsym} \, .
 \end{equation}
Hence, the  positive-$x$
and negative-$x$  components of the double 
distribution $\tilde F_{a}(x,y;t) $
can be treated as nonforward generalizations of 
quark and antiquark densities, respectively.
If we define the ``untilded'' DDs by 
\begin{equation} 
F_{a}(x,y;t) = \tilde F_{a}(x,y;t)|_{x>0}  
\ ; \ 
F_{\bar a}(x,y;t) = - \tilde F_{a}(-x,1-y;t)|_{x<0}   \, ,
\label{abara}   \end{equation}
then $x$ is always positive and 
 the reduction formulas  have  the same form
\begin{equation} 
\int_0^{1-x} \, F_{a,\bar a}(x,y;t=0)|_{x \neq 0} \, dy= 
 f_{a,\bar a}(x) 
 \label{34} \end{equation}
in both cases. The new antiquark distributions  
also ``live'' on the triangle
$0 \leq x,y \leq 1, \, 0 \leq x +y \leq 1$.
Taking $z$ in the lightcone ``minus''  direction,
we arrive at  the   parton interpretation 
of  functions $ F_{a, \bar a} (x,y;t )$ as probability amplitudes for an 
outgoing  parton to carry the fractions $xp^+$
and $yr^+$ of the external 
momenta $r$ and $p$. 
The  double distributions 
$F(x,y;t)$  are universal functions 
describing the flux of $p^+$ and $r^+$ 
independently of the ratio $r^+/p^+$. 
Note, that  extraction  of two separate
components $F_a(x,y;t)$ and  $F_{\bar a}(x,y;t)$ 
from the quark DD  $\tilde F_a(x,y;t)$ as its 
positive-$x$ and negative-$x$ parts  is unambiguous.

Taking the $O(z)$ term of the Taylor expansion  gives the 
sum rules 
\begin{equation} \int_0^1  dx\, \int_0^{1-x}  
 \left [F^a(x,y;t) - F^{\bar a}(x,y;t) \right ] \, dy
 =F_1^a (t) \, , \label{2}
\end{equation} 
\begin{equation} \int_0^1  dx\, \int_0^{1-x}  
 \left [K^a(x,y;t) - K^{\bar a}(x,y;t) \right ] \, dy
 =F_2^a (t) \, , \label{222}
\end{equation} 

\newpage 

\noindent relating the double distributions $ F_a(x,y;t)$,  
$ K_a(x,y;t)$ to 
the $a$-flavor components of the Dirac and Pauli
form factors:
\begin{equation}
\sum_a e_a F^a_1(t) = F_1(t)  \ \  , \  \  
\sum_a e_a F^a_2(t) = F_2(t) \,  ,
\end{equation}
 respectively.

A  common element of the reduction formulas given above 
is  an integration over $y$.  Hence, it is convenient 
to introduce   intermediate functions
\begin{equation}
{\cal F}^a(x;t) = \int_0^{1-x}  
 F^a(x,y;t)  \, dy  \ \  ;  \   \  {\cal K}^a(x;t) = \int_0^{1-x}  
 K^a(x,y;t)  \, dy \, .  \label{3a} 
\end{equation}
 They satisfy the reduction formulas
\begin{equation}  {\cal  F}^a(x;t=0) 
= f_a(x)  \  \  ;  \  \   \sum_a e_a \int_0^1 
\left [ {\cal F}^a(x;t) -   
 {\cal F}^{\bar a} (x;t)  \right ] 
 \,  dx  
 =F_1(t) \,  \label{3b} \end{equation}
 \begin{equation} 
\sum_a e_a \int_0^1   
 \left [ {\cal K}^a (x;t) - 
{\cal K}^{\bar a} (x;t)  \right ] 
 \,  dx  =F_2(t) \, , \label{3c}
\end{equation}
which show that these functions are 
the simplest  hybrids of the usual parton densities 
and form factors.
For this reason, one can   call them 
 {\it nonforward parton densities} (NDs) \cite{realco}. 
 
 The spin-flip terms disappear only if  $r=0$.  
 In the  weaker  $r^2 \equiv t =0$  limit,
they survive,
$e.g.,$ $F^a_2(0)= \kappa^a$ is the $a$-flavor 
contribution to the nucleon anomalous magnetic moment.
In other words,  the $t=0$ limit of the  ``magnetic'' 
NDs  exists:
${\cal K}^a (x;t=0) \equiv k_a(x)$, 
and the integral
\begin{equation}
\sum_a e_a \int_0^1   
 \left [ k_a(x) - k_{\bar a}(x) 
  \right ] 
 \,  dx  = \kappa_p \, \label{4}
\end{equation}
gives the anomalous magnetic moment of the proton. 
The  knowledge of the $x$-moment  of  $k_a(x)$'s
is needed to determine the contribution of the
quark orbital angular momentum to the proton spin \cite{ji}.
Since  the  $K$-type DDs  are always accompanied 
by the $r_{\mu} = p_{\mu}-p'_{\mu}$ factor,  they
 are invisible in deep inelastic
scattering and other inclusive processes 
related to strictly forward $r=0$ 
 matrix elements.

There are also parton-helicity 
sensitive double distributions $ G^{ a}(x,y;t)$ and  
$ P^{ a} (x,y;t)$. 
The first one  
reduces to the usual spin-dependent 
densities $\Delta f_a(x)$ in the 
$r=0$ limit and gives the axial 
form factor $F_A(t)$ after the $x,y$-integration.
The second one involves hadron helicity flip
and is  related to the pseudoscalar form factor 
 $F_P(t)$.

It is worth mentioning here that 
for a massive target  (nucleons in our case)
there is a kinematic restriction \cite{afs} 
\begin{equation}
-t > \zeta^2 M^2/\bar \zeta. 
\end{equation}
Hence,  for fixed $\zeta$, the formal limit $t \to 0$
is not  physically reachable. 
However, many results (evolution equations being the
most important example)
obtained in the formal $t =0$, $M=0$  
limit are still applicable.

\begin{figure}[t]
\mbox{
   \epsfxsize=5cm
 \epsfysize=5cm
 \hspace{4cm}   
\epsffile{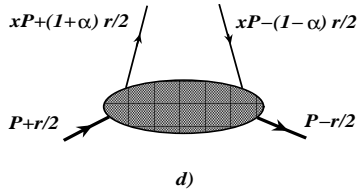} 
   }
  \vspace{-2cm}
{\caption{\label{fg:alpha} Parton picture
in terms of $\alpha$-DDs. 
   }}
\end{figure}

To make the description more symmetric 
with respect to the  initial and final hadron momenta,  
we can  treat nonforward matrix elements as 
functions of $(Pz)$ and $(rz)$, where $P=(p+p')/2$
is the average hadron momentum. 
The relevant  double distributions 
$\tilde f_a(x,\alpha\,;t)$ [which we 
will call $\alpha$-DDs to distinguish them
from $y$-DDs $F(x,y;t)$] are defined by  
\begin{eqnarray}
&&  \langle p' |  \bar \psi_a (-z/2) \hat z
\psi_a (z/2) |p \rangle \,\label{17} \\  = 
&& \bar u(p') \hat z u(p) \, \int_{-1}^1 \, dx \int_{-1+|x|}^{1-|x|} 
 e^{-ix(Pz)-i\alpha (rz)/2}  \, \tilde f_a(x,\alpha;t) 
 \,
d\alpha  \,   
 + O(r) \  {\rm terms} \  . \nonumber 
  \end{eqnarray}
The support area for $\tilde f_a(x,\alpha;t)$ is shown 
in Fig.\ref{fg:support}b.
Again, the usual forward densities 
$f_a(x)$ and $f_{\bar a} (x)$ 
are given by integrating 
$\tilde f_a(x,\alpha\,;\,t=0)$  over vertical lines
$x = {\rm const} $ for $x>0$ and $x<0$, respectively.
Hence, we can  split $\tilde f_a(x,\alpha\, ;\,t)$ 
into  three components 
\begin{equation} 
\tilde f_a(x,\alpha\,;\,t) = f_a(x,\alpha\,; \,t)\, \theta (x>0) - 
f_{\bar a} (-x, - \alpha\,; \,t)\, 
\theta (x<0) + f_M (x, \alpha\,; \,t) \, ,
\label{fff} \end{equation}
where $f_M (x, \alpha\,; \,t)$ is a singular  term
with support at $x=0$ only.  Due  to hermiticity
and time-reversal invariance properties of
nonforward matrix elements,  the $\alpha$-DDs
are even functions of $\alpha$: 
$$
\tilde f_a(x,\alpha;t) = \tilde f_a(x, - \alpha;t) \, . 
$$

\newpage 

\noindent  For our original $y$-DDs $F_{a, \bar a} (x,y;t)$, this 
corresponds to symmetry with respect to  the interchange 
$y \leftrightarrow 1-x-y$ (``Munich'' symmetry, 
established in Ref. \cite{lech}). In
particular, the  functions  $\varphi_M^{\pm}(y)$ for singular contributions
$F_M^{\pm}(x,y;t)$ are  symmetric  
$\varphi_M^{\pm}(y) = \varphi_M^{\pm}(1-y) $
both for  $x$-even and $x$-odd parts.
The  $a$-quark contribution into the 
 flavor-singlet 
operator
$$
{\cal O}_{ a}^S(-z/2,z/2) =  
 \frac{i}{2}
[ \bar \psi_a(-z/2) 
\hat z E(-z/2,z/2;A)  \psi_a(z/2)
- \{ z \to -z \}  ]   
  $$
can be  parametrized either by  $y$-DDs $\tilde F_a^S (x,y;t)$ or  
by $\alpha$-DDs $\tilde f_a^S(x, \alpha \, ; \, t)$  
\begin{eqnarray} 
&& \langle \,  p',s' \,  |     \,  \, {\cal O}_{ a}^S(-z/2,z/2)\, 
\,  |     \,  \, p,s \rangle \,  |     \, _{z^2=0}\label{311Q}  \\  &&  
= 
 \bar u(p',s')  \hat z  u(p,s)   
\int_0^1  dx \int_0^{1-x}   \, 
  \frac1{2} \left( e^{-ix(pz)-i(y-1/2)(r z)} \right. \nonumber \\ 
&& \left. \hspace{2cm} - e^{ix(pz)+i (y-1/2) (r z)}\right ) 
\,  F_a^S(x,y;t)  
  \, dy  + O(r) \ {\rm terms} \nonumber 
\\  &&  
= \bar u(p',s')  \hat z 
 u(p,s) \,  \int_{-1}^1  dx  \int_{-1+|x|}^{1-|x|}   
    e^{-ix(Pz)-i\alpha (r z)/2} 
 \tilde f_a^S(x,\alpha \, ;  t)  
 \, d\alpha  + O(r) \ 
.\nonumber 
\end{eqnarray} 
In the second and third lines here we have used the fact that 
  positive-$x$ and negative-$x$  parts 
  in this case 
are described by the  same untilded function  
$$
  F_a^S(x,y;t)|_{x \neq 0}   =  
 F_a(x,y;t) + F_{\bar a} (x,y;t)  .
$$
The  $\alpha$-DDs $\tilde f_a^S(x, \alpha \, ; \, t)$
are even functions of   $\alpha$ and, according to
Eq. (\ref{311Q}),    odd functions of $x$:
\begin{equation}
  \tilde f^{S}_a (x,\alpha ;t)   =   
  \{ f_a(|x|,|\alpha | ;t) + f_{\bar a} (|x|,|\alpha| ;t) \} \, 
  {\rm sign} (x) +f_M^S(x,\alpha ;t) \ .
\label{singlet} \end{equation} 
Finally, the valence 
 quark functions $\tilde f_a^{V}(x, \alpha \, ; \, t)$
related to the operators 
$$  {\cal O}^V_a (-z/2,z/2) = \frac12 [\bar \psi_a(-z/2) 
\hat z  E(-z/2,z/2;A)   \psi_a(z/2)
+  \{ z \to -z \} ]$$
are even functions of both $\alpha$  and $x$:
\begin{equation}
  \tilde f^{V}_a (x,\alpha ;t)   =   
  f_a(|x|,|\alpha| ;t) - f_{\bar a} (|x|,|\alpha|;t) + f_M^V(x,\alpha ;t) \ .
\label{valence} \end{equation}

\newpage

\section{Models for double and skewed distributions}

The  reduction formulas and   interpretation of
the  $x$-variable  as the  fraction of 
  $p$ (or $P$) momentum 
suggest that the   profile of $F(x,y)$ (or $f(x,\alpha)$)  
in  $x$-direction is basically determined by the shape 
of $f(x)$. 
On the other hand, the profile in  $y$ (or $\alpha$) direction  
characterizes the spread of the parton momentum induced by
the momentum transfer $r$. 
In particular, since 
the $\alpha$-DDs $\tilde f(x,\alpha)$ 
are even functions of $\alpha$,
it make sense to write 
\begin{equation}
\tilde f(x,\alpha) =  h(x,\alpha) \,  \tilde f(x)  \, ,  \label{65n}
 \end{equation}
 where $h(x,\alpha)$ is an even function of $\alpha$ 
 normalized by 
\begin{equation}
 \int_{-1+|x|}^{1-|x|} h(x,\alpha) \, d\alpha \, =1.
 \end{equation}
We may expect that 
the $\alpha$-profile of $h(x,\alpha)$  
is similar to that of a symmetric distribution amplitude (DA) 
$\varphi (\alpha)$.  Since $|\alpha| \leq 1- |x| $, to get a 
more complete  analogy
with DA's, 
it makes sense to rescale $\alpha$ as $\alpha = (1-|x|)  \beta$
introducing  the  variable $\beta$ with $x$-independent limits:
$-1 \leq \beta \leq 1$. 
 The simplest model is to assume 
that the profile in the $\beta$-direction is  
 a  universal function  $g(\beta)$ for all $x$. 
Possible simple choices for  $g(\beta)$ may be  $\delta(\beta)$
(no spread in $\beta$-direction),  $\frac34(1-\beta^2)$
(characteristic shape for asymptotic limit 
of nonsinglet quark distribution amplitudes), 
 $\frac{15}{16}(1-\beta^2)^2$
(asymptotic shape of gluon distribution amplitudes), etc.
In the variables $x,\alpha $, this gives   
\begin{eqnarray}
&& h^{(\infty)} (x,\alpha) =  \delta(\alpha) \,   \ , \
h^{(1)}(x,\alpha) = \frac{3}{4} 
\frac{ (1- |x|)^2 - \alpha^2}{(1-|x|)^3} \,   \ ,  \nonumber \\
&& h^{(2)}(x,\alpha) = \frac{15}{16} 
\frac{[(1- |x|)^2 - \alpha^2]^2}{(1-|x|)^5} \,   \  . \label{mod123} 
 \end{eqnarray}
 These models can be treated as specific
 cases of the general profile function 
 \begin{equation}
 h^{(b)}(x,\alpha) = \frac{\Gamma (2b+2)}{2^{2b+1} \Gamma^2 (b+1)}
\frac{[(1- |x|)^2 - \alpha^2]^b}{(1-|x|)^{2b+1}} \,  , \label{modn} 
 \end{equation}
whose width is governed by the parameter $b$.

\begin{figure}[htb]
%\hspace{-1cm} 
\mbox{
   \epsfxsize=11cm
 \epsfysize=6.5cm
  \epsffile{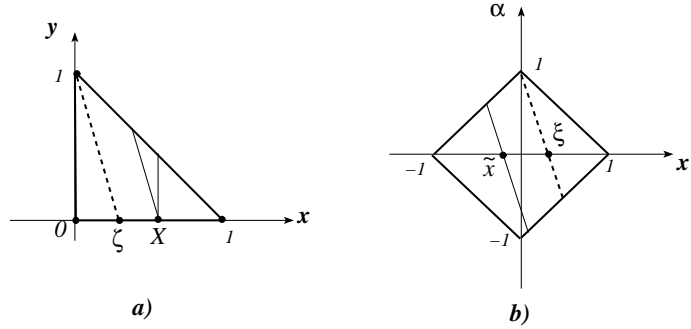}  }
  \vspace{-1cm}
{\caption{\label{fg:spds}    Integration lines 
for integrals relating SPDs and DDs. 
   }}
\end{figure}

The  coefficient of proportionality 
$\zeta = r^+/p^+$  (or $\xi = r^+/2P^+$)  between 
the plus components of the momentum transfer 
and initial (or  average) momentum specifies 
the {\it skewedness}  of the matrix elements.
The characteristic   feature  implied by
 representations for double distributions 
 [see, e.g., Eq.(\ref{31})]
 is the absence 
of the $\zeta$-dependence in 
the DDs   $F(x,y)$ and $\xi$-dependence in $f(x,\alpha)$.
An alternative way to parametrize 
nonforward matrix elements of light-cone operators
is to use   
 $\zeta$  (or $\xi$) and the {\it total } 
 momentum fractions  $X \equiv x+y \zeta$ 
(or $\tilde x \equiv  x +\xi \alpha$) 
  as  independent 
variables. 
Integrating each particular  double   distribution 
over $y$ gives  the nonforward  parton distributions  
\begin{eqnarray} && 
{\cal F}_{\zeta}^{i} (X) = \int_0^1 dx \int_0^{1-x} 
\, \delta (x+\zeta y -X) \, F_i(x,y) \, dy \label{71}  \\ &&
=
\theta(X \geq \zeta) 
 \int_0^{ \bar X / \bar \zeta } F_{i}(X-y \zeta,y) \, dy + 
 \theta(X \leq \zeta)  \int_0^{ X/\zeta} F_{i}
 (X-y \zeta,y) \, dy \,  , \nonumber
 \end{eqnarray}
where $\bar \zeta \equiv 1- \zeta$.
The two components of NFPDs correspond to positive
($X> \zeta$) and negative ($X< \zeta$)
values of the fraction $X' \equiv X - \zeta$ 
associated with the ``returning'' parton.
As explained in refs. \cite{compton,npd},
the second component  can be interpreted as the
probability amplitude for the initial hadron with momentum 
$p$ to split into the final hadron with momentum $(1-\zeta)p$
and the two-parton state with total momentum $r=\zeta p$
shared by the partons
in fractions $Yr$ and $(1-Y)r$, where $Y=X/\zeta$.

The relation between ``untilded'' NFPDs
and  DDs  can be     illustrated
on the ``DD-life triangle'' 
defined by $0 \leq x,y,x+y \leq 1$ (see Fig.\ref{fg:spds}a).
Specifically, to get ${\cal F}_{\zeta} (X)$,  one should 
integrate $F(x,y)$ over $y$ along a straight
line  $x=X- \zeta y$. Fixing some value of  $\zeta$,
one deals with  a set of parallel lines  intersecting  the $x$-axis
at $x=X$. The upper limit of the $y$-integration
is determined by intersection of this line
either with the line $x+y=1$ (this happens if 
$X > \zeta$)  or with the $y$-axis (if $X < \zeta$).
The line corresponding to $X=\zeta$
separates the triangle into two parts
generating the  two components of the
nonforward parton distribution.
  
  \newpage

In a similar way, we can write the relation 
between OFPDs 
$H(\tilde x,\xi;t)$  
 and the 
$\alpha$-DDs $\tilde f(x,\alpha;t)$ 
\begin{equation} 
  H (\tilde x,\xi; t)=  \int_{-1}^1 dx \int_{-1+|x|}^{1-|x|}  
\, \delta (x+ \xi \alpha - \tilde x) \, 
\tilde f (x,\alpha;t) \, d\alpha  \, . \label{710} 
 \end{equation} 
The  delta-function in Eq.(\ref{710}) specifies 
the line of  integration in the
$\{ x, \alpha \}$ plane. For definiteness,  we will assume below
that 
$\xi$ is positive.

Information contained in SPDs originates from two  
physically different sources: meson-exchange type contributions  
${\cal F}_{\zeta}^M(X)$ 
coming from the singular $x=0$ parts of DDs
and  the functions 
${\cal F}_{\zeta}^a(X)$,  ${\cal F}_{\zeta}^{\bar a}(X)$
 obtained by scanning the $x \neq 0$  parts of  
DDs $F^a(x,y)$, $F^{\bar a} (x,y)$. 
The support of exchange contributions is restricted 
to $0 \leq X \leq \zeta$. Up to rescaling, the function
${\cal F}_{\zeta}^M(X)$  has the same shape for all $\zeta$.
For any nonvanishing $X$, these exchange terms  become  invisible 
in the forward limit $\zeta \to 0$.  
On the other hand, the support of functions 
${\cal F}_{\zeta}^a(X)$,  ${\cal F}_{\zeta}^{\bar a}(X)$
in general covers the whole   $0\leq X \leq 1$ region. 
Furthermore, the forward limit of such  SPDs as 
${\cal F}_{\zeta}^{a, \bar a}(X)$ is   
rather well known  from inclusive measurements.  
Hence, information  contained in the usual (forward) densities
$f^a(x)$, $f^{\bar a}(x)$ can be used to 
restrict the models for 
${\cal F}_{\zeta}^a(X)$,  ${\cal F}_{\zeta}^{\bar a}(X)$.

Let us consider 
SPDs  constructed using    simple 
models of DDs specified above.
In particular,  the model
$f^{(\infty )}(x,\alpha) = \delta (\alpha)
f(x)$ (equivalent to $F^{(\infty)}(x,y) = \delta (y -\bar x/2)
f(x)$), gives the simplest    model 
  $H^{(\infty)}(\tilde x,\xi; t=0) = f(x)$ in which  OFPDs at $t=0$
   have no $\xi$-dependence.
For NFPDs this gives  
\begin{equation}
{\cal F}_{\zeta}^{(\infty)} (X) = \frac{\theta(X \geq \zeta/2)}{1-\zeta/2}
 f \left (\frac{X-\zeta/2}{1-\zeta/2} \right ) \, ,
 \label{model}
  \end{equation}
i.e.,  NFPDs for non-zero  $\zeta$ are obtained from 
the forward distribution $f(X)\equiv {\cal F}_{\zeta=0} (X)$  
 by   shift and rescaling.

\begin{figure}[htb]
\mbox{
   \epsfxsize=6cm
 \epsfysize=4cm
 \hspace{0cm}  
  \epsffile{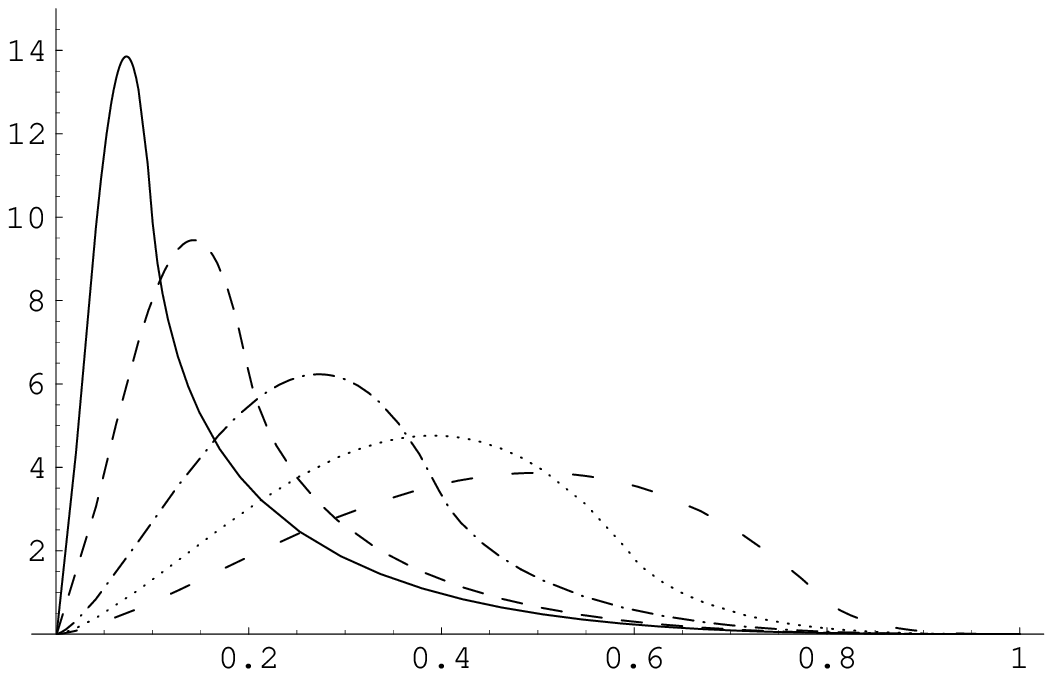}   \hspace{0cm}
   \epsfxsize=6cm
 \epsfysize=4cm
 \hspace{0cm}  
  \epsffile{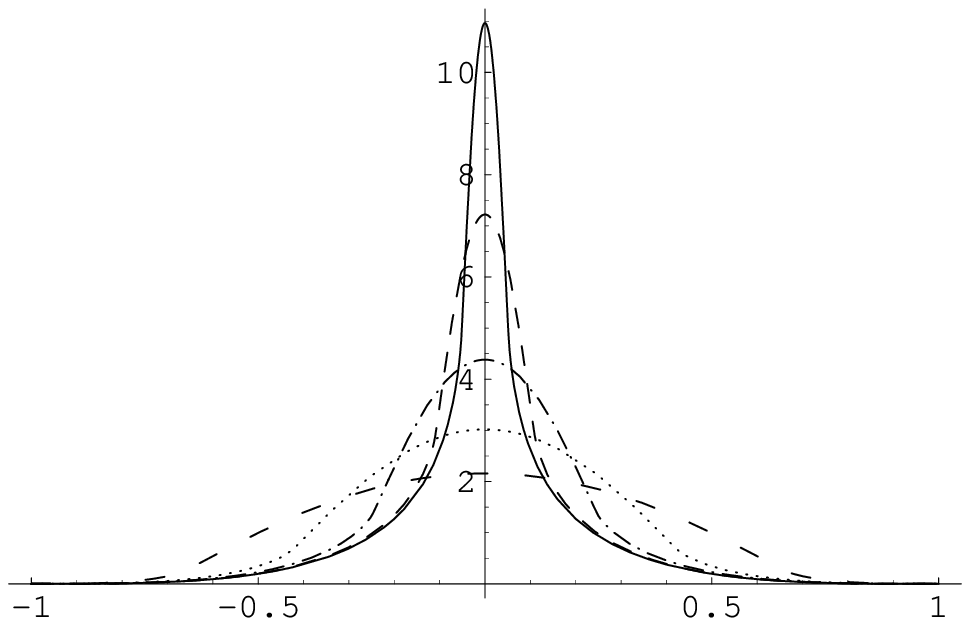}  }
{\caption{\label{q-at-diff-z}
Valence quark  distributions: untilded NFPDs $F^{q}_\zeta(x)$ (left)
and OFPDs $H ^1_V(\tilde x,\xi)$ (right) with $a= 0.5$ 
for several values
of $\zeta$, namely,  0.1, 0.2, 0.4, 0.6, 0.8 and corresponding values
of $\xi=\zeta / (2-\zeta)$. Lower curves
correspond to larger values of $\zeta$.}}
\end{figure}

In case of  the $b=1$ and $b=2$  models, simple  analytic results 
can be obtained only for some   explicit forms of  $f(x)$.
For the ``valence quark''-oriented ansatz $\tilde f^{(1)}(x,\alpha)$,
the following choice of a normalized distribution
\begin{equation}  f^{(1)}(x) = 
 \frac{\Gamma(5-a)}{6 \,  \Gamma(1-a)} \, 
  x^{-a} (1-x)^3  \label{74} \end{equation}
is   both  close 
to phenomenological   quark distributions
and   produces a simple expression
for the double distribution since the denominator
$(1-x)^3$ factor in Eq. (\ref{mod123}) is canceled.
As a result, the integral in Eq. (\ref{710})
is easily performed and   we get
\begin{eqnarray}
 H^{1 }_V(\tilde x, \xi)|_{|\tilde x| \geq \xi}  = \frac1{\xi^3} 
 \left ( 1- \frac{a}{4} \right ) 
 \biggl  \{  \biggl [ (2-a) \xi 
(1- \tilde x) (x_1^{2-a} + x_2^{2-a}) 
 \nonumber \\ 
  +
(\xi^2 -\tilde x)(x_1^{2-a} - x_2^{2-a})  \biggr ] \, \theta (\tilde x) 
+ ( \tilde x \to -\tilde x)  \biggr \}  \label{outs} 
\end{eqnarray}
for  $|\tilde x |\geq \xi$ and 
\begin{eqnarray}
H^{1 }_V (\tilde x, \xi)|_{|\tilde x| \leq \xi}  = \frac1{\xi^3} 
\left ( 1- \frac{a}{4} \right ) \biggl \{ x_1^{2-a}[(2-a) \xi (1- \tilde x) +
(\xi^2 -x)]  \nonumber \\ 
+ ( \tilde x \to -\tilde x) \biggr \} \label{middles}
\end{eqnarray}
in the middle $ -\xi \leq \tilde x \leq \xi$ region.
We use here  the notation $x_1=(\tilde x + \xi)/(1+\xi)$
  and 
$x_2=(\tilde x - \xi)/(1-\xi)$ \cite{jirev}.
To extend these expressions onto negative values of 
$\xi$, one should  substitute $\xi$ by $|\xi|$.
One can check, however, that no odd powers of $|\xi|$ 
would appear in the $\tilde x^N$ 
moments of $H^{1V }(\tilde x, \xi)$.
Furthermore, these expressions are explicitly non-analytic 
for $x = \pm \xi$. 
This  is true even if $a$ is integer.
Discontinuity at $x = \pm \xi$, however, appears only 
in the second derivative
of $H^{1V }(\tilde x, \xi)$, 
i.e., the model curves for $H^{1V }(\tilde x, \xi)$
look very smooth (see Fig.\ref{q-at-diff-z}).
The explicit expressions for NFPDs in this 
model were given in ref.\cite{ddee}. The relevant curves 
are also shown in Fig.\ref{q-at-diff-z}.

For   $a=0$, the 
$x>\xi$ part of OFPD has the same $x$-dependence 
as its forward limit, differing from it by an overall $\xi$-dependent 
factor only:
\begin{equation}
H^{1V }(\tilde x, \xi)|_{a=0} = 
4 \, \frac{(1-|\tilde x|)^3}{(1-\xi^2)^2} 
\, \theta (|\tilde x| \geq \xi) \, 
+ 2\, \frac{\xi +2 -3 \tilde x^2/\xi}{(1+\xi)^2} \, 
 \theta (|\tilde x| \leq \xi)
\, .    \label{(1-x)^3}
\end{equation} 
The  $(1-|\tilde x|)^3$ behaviour can be 
trivially continued into the $|\tilde x| < \xi$ 
region. However, the actual behaviour
of $H^{1V }(\tilde x, \xi)|_{a=0}$ in this region 
is given by a different function.
In other words, $ H^{1V }(\tilde x, \xi)|_{a=0}$
can be represented as a sum of a function analytic at 
border points and a contribution  whose support 
is restricted by $|\tilde x| \leq \xi$.  
It should be emphasized that despite its DA-like 
appearance, this contribution 
should not be treated as an exchange-type term. 
It is generated by regular $x \neq 0$ part of DD,
and, unlike $\varphi (\tilde x / \xi)/\xi$ functions
 changes its shape with $\xi$ and  becomes very small  for small $\xi$.

\begin{figure}[htb]
\mbox{
   \epsfxsize=8cm
 \epsfysize=5cm
 \hspace{2cm}  
  \epsffile{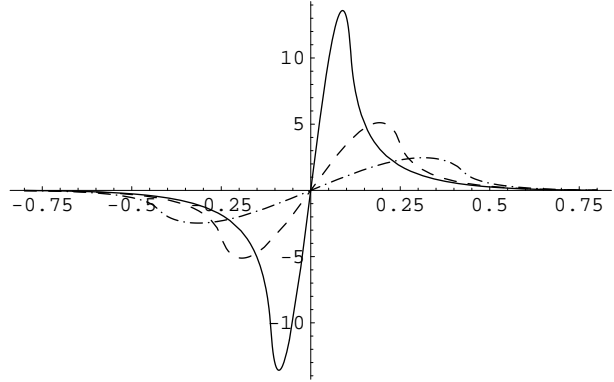}  }
{\caption{\label{qs-at-diff-z}
Singlet quark distribution $H ^1_S(\tilde x,\xi)$  for 
several $\xi$ values
0.1, 0.25, 0.4. Lower curves
correspond to larger values of $\xi$. Forward distribution
is modeled by \mbox{$(1-x)^3/x$.}}}
\end{figure}

For  the singlet quark distribution, the  $\alpha$-DDs
$\tilde f^S( x, \alpha)$ should be odd functions 
of $x$. Still, we can use  the model like (\ref{74}) for 
the $x>0$ part, but 
take $\tilde f^S( x, \alpha)|_{x \neq 0} 
= A \, f^{(1)}( |x|, \alpha)\, {\rm sign} (x)$. 
Note, that the integral (\ref{710})  producing $H^S(\tilde x, \xi)$ 
in the $|\tilde x| \leq \xi$ region
would diverge  for  $\alpha \to \tilde x /\xi$  
 if $a \geq  1$, which is the usual case 
 for standard parametrizations of singlet quark 
 distributions for sufficiently large $Q^2$. 
However, due to the antisymmetry of  $\tilde f^S( x, \alpha)$
wrt  $x \to -x$ and its symmetry wrt $\alpha \to -\alpha$,
the singularity at 
$\alpha = \tilde x /\xi$ can be  integrated  
using the principal value 
prescription   which in this case 
produces the $x\to -x$ antisymmetric version 
of Eqs.(\ref{outs}) and (\ref{middles}). For $a=0$, 
its middle part  reduces to 
\begin{equation}
H^{1S}(\tilde x, \xi)|_{|\tilde x| \leq \xi, a=0} = 
 2x\, \frac{3 \xi^2 -2 x^2 \xi - x^2}{\xi^3 (1+\xi)^2} 
  \,  .
\end{equation}
 The shape of singlet SPDs in this model is shown in Fig.
\ref{qs-at-diff-z}
 
\section{ SPDs and deeply virtual Compton scattering}

In the lowest order,  the DVCS amplitude 
$T^{\mu \nu} (p,q,q')$  is given by two
handbag diagrams.  In particular, the invariant  amplitude 
containing the ${\cal F}$ functions is given by
\begin{eqnarray} 
T_F(p,q,q') = \sum_a e_a^2 
 \int_0^{1} 
\left  [ \frac1{X-\zeta +i\epsilon}
+ \frac1{X- i \epsilon} \right ]  {\cal F}^{a+ \bar a}_{\zeta}(X;t)
\, dX \,  . 
\label{123}  \end{eqnarray} 
 An important  feature of the DVCS amplitude is that 
for large $Q^2$ and fixed $t$ it depends only on the ratio 
$Q^2/2 (pq) \equiv x_{Bj}
=\zeta$:  DVCS is an  {\it exclusive} process 
exhibiting the Bjorken  scaling. 
Note that the imaginary part of the DVCS amplitude is proportional
to ${\cal F}^{a+ \bar a}_{\zeta}(\zeta;t)$.
In this function, the parameter $\zeta$ appears twice:
first as the skewedness of the process 
and then as the fraction $X=\zeta $ 
at which the imaginary part is generated. 

One may ask which  $Q^2$   are large enough to ensure 
the dominance of the lowest-twist  handbag contribution. 
In DIS, approximate Bjorken scaling  starts at $Q^2 \sim 2 \, $GeV$^2$.
Another example  is given by the exclusive
process  $\gamma(q_1) \gamma^*(q_2) \to \pi^0$ studied on 
$e^+ e^-$ colliders.  If one of the photons  is highly virtual
$q_1^2 = -Q^2$ while another is (almost) real $q_2^2 \sim 0$,
the process is kinematically
similar  to DVCS.  
In the leading order, the $F_{\gamma \gamma^* \pi^0}(Q^2)$ 
transition form factor  is  given by a handbag diagram again. 
The recent measurements by CLEO \cite{cleo} 
show that the pQCD prediction $F_{\gamma \gamma^* \pi^0}(Q^2) \sim 1/Q^2$ 
again works starting  from $Q^2 \sim 2 \, $GeV$^2$.
The $\gamma \gamma^* \pi^0$ vertex  (for a virtual pion) can be also 
measured on a fixed-target machine  like CEBAF 
in which case it is just a part
of the DVCS amplitude  corresponding to the 4th skewed distribution
${\cal P}_{\zeta}(X,t)$ (which is related to the pseudoscalar 
form factor  $G_P(t)$ of  the nucleon).
Hence, CLEO data  give an evidence that DVCS  may be 
handbag-dominated for $Q^2$ as low as 2\,GeV$^2$.

The main problem for studying DVCS is the contamination 
by the Bethe-Heitler process in which the final photon
is emitted from the initial or final electron.
The Bethe-Heitler amplitude is enhanced at small $t$.
On the other hand, the virtual photon flux 
for fixed $Q^2$ and $x_{Bj}$ increases when 
the electron beam energy increases.  Hence,  the  energy  upgrade 
 would make the DVCS studies  at  Jefferson Lab 
more feasible. Experimental aspects  of virtual
Compton scattering studies at Jefferson Lab are discussed
in Refs. \cite{vanguigui}.

The skewed parton distributions can be also 
measured  in hard meson electroproduction processes
\cite{npd,cfs,lech,vanguigui}.   
The leading-twist pQCD contribution in this case
involves a  one-gluon exchange, which means that the hard
subprocess  is suppressed by  $\alpha_s/\pi \sim 0.1$ 
factor. The competing soft mechanism corresponds to
a triple overlap of hadronic wave functions and 
has   a relative  suppression  $M^2/Q^2$ by a power of  
$Q^2$,  with $M^2 \sim 1$\, GeV$^2$  being a characteristic
hadronic scale. Hence, to clearly see the 
one-gluon-exchange signal one needs $Q^2$ above 10\,GeV$^2$.
Numerical pQCD-based estimates 
and comparison of DVCS and hard meson electroprodution 
cross sections 
can be found in  Ref. \cite{vanguigui}.

\section{SPD enhancement factor}

 The imaginary part of  hard exclusive 
 meson electroproduction amplitude is determined by
 the skewed distributions at 
 the border point. For this reason, the magnitude 
 of  ${\cal F}_{\zeta} (\zeta)$ [or $H(\xi, \xi)$]
   and its relation to the  forward densities
 $f(x)$   has a practical interest. 
 This example also gives a  possibility to study  
 the sensititivity of the results to the choice of the
 profile function.  
 Assuming   the infinitely narrow weight 
 $\rho(\alpha) = \delta (\alpha)$,
 we have ${\cal F}_{\zeta} (X ) = f(X-\zeta/2) + \ldots $ 
 and $H(x,\xi) = f(x)$.
 Hence, both ${\cal F}_{\zeta} (\zeta)$ and  $H(\xi, \xi)$
 are given by $f(x_{Bj}/2)$ since $\zeta = x_{Bj}$
 and $\xi =x_{Bj}/2 +\ldots$.  Since the argument
 of $f(x)$ is twice smaller than in deep inelastic scattering,
this results in an enhancement factor. In particular, if 
  $f(x) \sim x^{-a}$ for small  $x$, the ratio
  ${\cal R} (\zeta ) \equiv {\cal F}_{\zeta} (\zeta ) /f(\zeta )$ is 
  $2^a$. 
 The use of a wider profile  function $\rho (\alpha)$ produces further
 enhancement. For example, taking the normalized profile 
 function 
 \begin{equation}
 \rho_b (\alpha)\equiv  \frac{\Gamma (b+3/2) }{ \Gamma (1/2) 
 \Gamma (b+1) } (1- \alpha^2)^b = 
  \frac{\Gamma (2b+2)}{2^{2b+1} \Gamma^2 (b+1)}
 (1-\alpha^2)^b \label{rhon}
 \end{equation} 
 and
$f(x) \sim x^{-a}$ we get 
 \begin{equation}{\cal R}^{(b)} (\zeta) \equiv 
 \frac{{\cal F}_{\zeta}^{(b)}   (\zeta ) }{f(\zeta )}
  = \frac{\Gamma (2b+2)\Gamma (b-a +1)}  
  {\Gamma (2b-a+2)\Gamma (b +1)}
 \end{equation}
 which is larger than $2^a$ for any finite $b$ and $0< a <2$.
 The $2^a$ enhancement appears as the $b \to
  \infty$ limit of Eq.(\ref{rhon}).
 For small integer $b$, Eq.(\ref{rhon})  reduces
 to simple formulas obtained in refs. \cite{ddee,sssdd}.
 For $b=1$, we have 
 \begin{equation}
 \frac{{\cal F}_{\zeta}^{(b=1)}   (\zeta ) }{f(\zeta )}
  = \frac1{(1-a/2)(1-a/3)}  \, ,
  \label{rho1} 
 \end{equation}
 which gives the factor of 3 for the  enhancement if $a=1$.
 For $b=2$, the ratio (\ref{rhon}) becomes
 \begin{equation}
 \frac{{\cal F}_{\zeta}^{(b=2)}   (\zeta ) }{f(\zeta )}
  = \frac1{(1-a/3)(1-a/4)(1-a/5)}  \, ,
  \label{rho2} 
 \end{equation}
  producing a smaller enhancement factor $5/2$ for $a=1$. 
 Calculating the enhancement factors,
one should remember that the gluon SPD ${\cal F}_{\zeta}(X)$ 
reduces to $Xf_g(X)$ in the $\zeta =0$ limit. 
Hence, to get the enhancement factor corresponding to  the 
$ f_g(x) \sim x^{-\lambda}$  small-$x$ behavior of the forward 
gluon density, one should take $a= \lambda -1$ in Eq.(\ref{rhon}),
i.e., despite the fact that the $1/x$ behavior of the 
singlet quark distribution gives the factor of 3
for the ${\cal R}^{(1)} (\zeta)$ ratio, the same shape of 
the gluon distribution 
results in no enhancement.

 Due to evolution, the effective parameter $a$ 
characterizing the small-$x$ behavior 
of the forward distribution 
is an increasing function of $Q^2$.
As a result, for fixed $b$, the ${{\cal R}^{(b)} 
  (\zeta ) }$  
ratio  increases with $Q^2$.  
In general, the  profile of $\tilde f (\tilde x,  \alpha  )$
 in the $\alpha$-direction
is also  affected by the pQCD  evolution. 
In particular, in ref. \cite{ddee} it was shown that
  if one takes an ansatz  corresponding 
  to an extremely  asymmetric profile function $\rho (\alpha) 
  \sim  \delta (1+\alpha)$,  the shift 
  of the profile function to a more symmetric 
  shape  is clearly visible in the evolution 
  of the relevant SPD. 
  Recently, it was demonstrated  \cite{sgmr,murad} that 
  evolution to sufficiently large $Q^2$ 
  enforces a direct relation $b=a$ between the 
  parameter $a$ characterizing the small-$x$ 
  behavior of DDs and the parameter $b$ governing the shape 
of their $\alpha$ profile.
This gives 
\begin{equation}{\cal R}^{(b=a)} (\zeta) 
  = \frac{\Gamma (2a+2)}  
  {\Gamma (a+2)\Gamma (a +1)}
 \end{equation}
for the ${\cal R}(\zeta)$ ratio. 
For $a=1$, e.g., the SPD enhancement factor
in this case equals 3.

\section{Compton scattering} 

\subsection {General Compton amplitude}

The Compton scattering in its various versions 
provides a unique tool for studying 
 hadronic structure.
The  Compton amplitude probes the hadrons through a 
 coupling of two electromagnetic currents
 and in this aspect it  can be considered 
as a generalization of hadronic form factors.  
In QCD, the photons interact with the 
quarks of a hadron through 
a vertex which, in the lowest
 approximation,  has a   pointlike structure.
However, in the soft regime, 
strong interactions  produce large 
corrections uncalculable within the
perturbative QCD framework.
To take advantage of  the basic pointlike structure of the
photon-quark coupling and the asymptotic
freedom feature of QCD, one should choose 
a specific kinematics  in which the behavior
of the relevant amplitude is dominated by 
short (or, being more precise, lightlike) distances.
The general feature of all such types of kinematics is the 
presence of a large momentum transfer.
For  Compton amplitudes, there are several
situations when large momentum transfer induces
dominance of configurations 
involving lightlike distances:  \\ 
$i)$ both photons are far off-shell 
and have  equal spacelike virtuality:
 virtual forward Compton amplitude,
 its imaginary part determines structure
 functions of deep inelastic scattering (DIS); \\ 
 $ii)$ initial photon is highly virtual,
 the final one is real and the momentum transfer 
 to the hadron is small: deeply virtual 
 Compton scattering (DVCS) amplitude; \\ 
 $iii)$ both photons  are real but the 
 momentum transfer
 is large:  wide-angle Compton 
 scattering (WACS) amplitude.

 Our main statement made in ref. \cite{realco} 
 is that, at accessible momentum transfers
 $|t| < 10$ GeV$^2$, the  WACS amplitude  is dominated 
 by  handbag diagrams, just like  in  DIS and DVCS.
 In the most general case, the nonperturbative 
 part of the handbag contribution
 is described by  double distributions (DDs) 
 $F(x,y;t),   G(x,y;t)$, etc.,  which can be 
 related  to the usual parton 
 densities $f(x)$, $\Delta f(x)$ and nucleon form factors
 like $F_1(t),G_A(t)$. 
 Among the arguments of DDs,   $x$ is 
  the fraction of the initial hadron momentum carried 
 by the active parton and $y$ is the  fraction   
 of the momentum transfer $r$.
The description of WACS amplitude simplifies 
when one can neglect the $y$-dependence 
 of the hard part and integrate out
    the $y$-dependence
 of the double distributions. In that case,
 the long-distance dynamics is described by nonforward
 parton densities (NDs)  ${\cal F}(x;t), {\cal G}(x;t),$ etc. 
 The latter 
 can be interpreted as the usual parton densities $f(x)$ 
 supplemented by a form factor type $t$-dependence.  
 We  proposed in \cite{realco}    a simple model 
 for the relevant NDs 
 which both satisfies the relation between  ${\cal F}(x;t)$ and
   usual parton densities $f(x)$ and  produces 
   a good description of 
  the  $F_1(t)$ form factor up to $t \sim  - 10$ GeV$^2$. 
  We have used this model to calculate  the WACS amplitude
  and obtained the results which are 
  rather close to existing data.  

\subsection{Deep inelastic scattering}

The forward virtual  Compton amplitude whose
imaginary part gives structure functions
of deep inelastic scattering (see, e.g., \cite{feynman})
is the classic example  of a 
light cone dominated Compton amplitude.
 In this case,
the ``final'' photon  has momentum $q'=q$ coinciding with
that of the initial one. The momenta $p,p'$ of the initial
and final hadrons also coincide. 
The total cm energy of the photon-hadron system
$s= (p+q)^2$ should be above resonance region,
and the Bjorken ratio $x_{Bj} = Q^2/2(pq)$ is finite.
The light cone   dominance is secured by high
virtuality of the photons: $-q^2 \equiv Q^2 > 1 $ GeV$^2$.
In the large-$Q^2$ limit,  the leading  contribution 
in the lowest $\alpha_s$  order 
is given by  handbag diagrams in which the 
perturbatively calculable hard quark
propagator is convoluted with parton distribution
functions $f_a(x)$ ($a=u,d,s, \ldots$) 
which describe/parametrize nonperturbative
information about hadronic structure.

\subsection{Deeply virtual Compton scattering}

The condition that both photons are highly virtual
may be relaxed by taking 
a  real photon in the final state.
Keeping the momentum transfer $t\equiv (p-p')^2$ 
to the hadron  as small as possible,
one arrives at  kinematics of the deeply virtual 
Compton scattering (DVCS) the  importance of which 
was recently emphasized by X. Ji \cite{ji}
 (see also \cite{compton}). 
 Having   large  virtuality $Q^2$  
of the initial photon is sufficient to  
guarantee that in the Bjorken limit 
the leading power  contributions in $1/Q^2$ 
are generated  by the strongest  light cone    
singularities \cite{ji,npd,jios2,cofre},  with   
the handbag diagrams being the  
starting point of the $\alpha_s$
expansion.  
The most important contribution to the 
DVCS amplitude is given  by a convolution
of a  hard quark propagator and 
a nonperturbative function describing 
long-distance dynamics, which in  the most general case is 
given by    double    distributions 
(DDs) $F(x,y;t), G(x,y;t), \ldots $ \cite{compton,npd}.

\begin{figure}[t]
\mbox{
   \epsfxsize=12cm
 \epsfysize=4.5cm
 %\hspace{4cm}  
  \epsffile{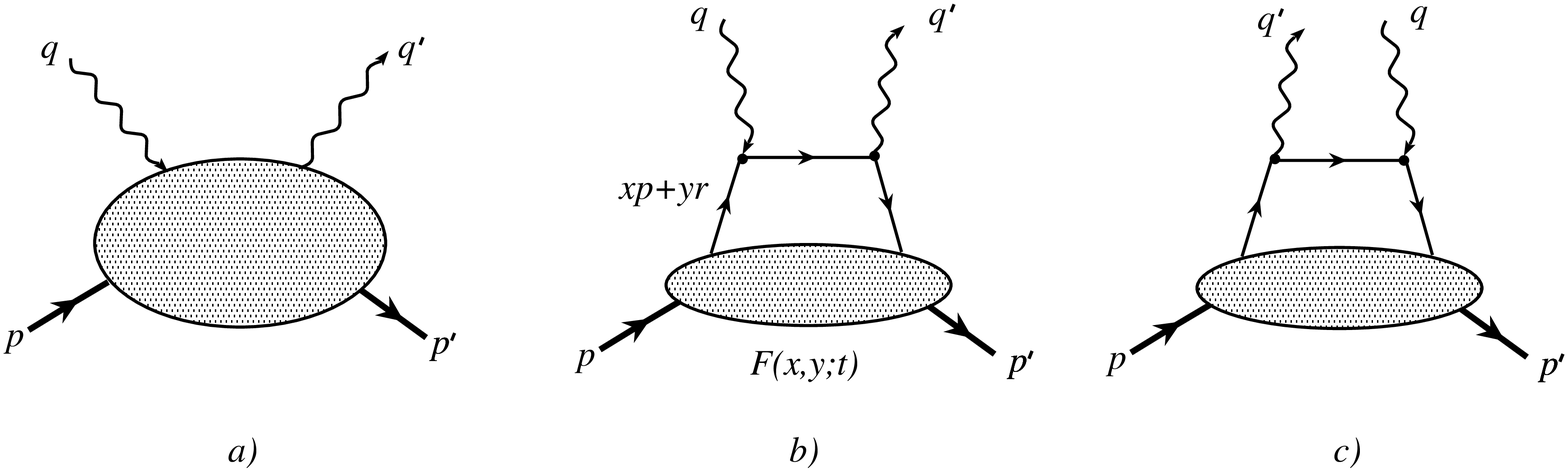}  }
  %\vspace{0.5cm}
{\caption{\label{fig:cohan} $a)$ General Compton amplitude;
$b)$ $s$-channel handbag diagram;  $c)$ $u$-channel handbag diagram.
   }}
\end{figure} 

In the DVCS kinematics,   $|t|$ is assumed to be small compared to 
$Q^2$, and for this reason 
the $t$- and $m_p^2$-dependence of the short-distance amplitude 
in refs. \cite{ji,compton,npd,jirev} was neglected\footnote{
One should not think that  such a dependence 
is necessarily a higher
twist effect: the lowest twist contribution
has a calculable dependence on $t$ and $m_p^2$ 
analogous to the Nachtmann-Georgi-Politzer 
 $O(m_p^2/Q^2)$ target mass corrections in DIS
\cite{nachtmann,geopol}.}. 
This is equivalent to approximating the 
active parton momentum $k$ by its plus component alone:
$k \to  xp^+ + yr^+$.

\section{Modeling NDs}

Our final  goal in the present paper is to  get an estimate  of 
  the handbag  contributions 
for the large-$t$ real Compton scattering.
 Since the initial   photon in that case is also     real: 
 $Q^2=0$ (and hence $x_{Bj}=0$), it is natural to expect that 
 the nonperturbative functions which appear in WACS 
 correspond   to the $\zeta = 0$ limit of 
 the  skewed  parton distributions\footnote{Provided 
 that one can neglect the $t$-dependence of the hard 
 part.} 
 ${\cal F}^{ a}_{\zeta}(x;t)$.
 It is easy to see from Eq.(\ref{3a})
 that in this limit
 the SPDs reduce to the  nonforward parton 
densities ${\cal F}^a(x;t)$
introduced above:  
\begin{equation} {\cal F}_{\zeta=0}^a(x;t) = 
{\cal F}^a(x;t) \, .\end{equation}
 Note that NDs depend on  
 ``only   two''  variables $x$ and $t$,
 with this dependence  
   constrained by reduction
 formulas (\ref{3b}),(\ref{3c}).  
 Furthermore,    it is possible 
to give an interpretation of nonforward  densities  
in terms of the  light-cone wave functions.

Consider for simplicity 
a two-body bound state  whose 
lowest Fock component is described by a light cone   
wave function $\Psi(x,k_{\perp})$.
Choosing a frame where the momentum transfer $r$  is purely 
transverse  $r=  r_{\perp}$, we can write  
the two-body contribution into the form factor 
as \cite{bhl}
\begin{equation}
F^{(tb)} (t) = \int_0^1 \,  dx \, \int \, 
\Psi^* ( x, k_{\perp}+\bar x r_{\perp}) \,  
\Psi (x, k_{\perp}) 
\, {{d^2 k_{\perp}}\over{16 \pi^3}}  \, , 
\end{equation}  
\begin{figure}[htb]
\mbox{
   \epsfxsize=12cm
 \epsfysize=5cm
 %\hspace{3cm}  
  \epsffile{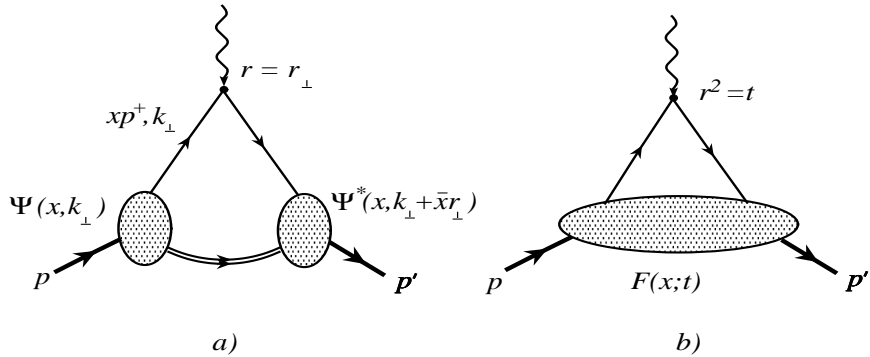}  }
  %\vspace{0.5cm}
{\caption{\label{fig:psifor} $a)$ Structure of the 
effective two-body contribution 
to form factor in the light cone formalism.
$b)$ Form factor as an $x$-integral  of  nonforward parton densities.
   }}
\end{figure}
\noindent where $\bar x \equiv 1-x$. 
  Comparing this expression with the reduction formula 
  (\ref{3b}), we conclude that 
 \begin{equation} 
 {\cal F}^{(tb)} (x,t) = \int  \, 
 \Psi^* ( x, k_{\perp}+ \bar x r_{\perp}) \,
\Psi (x, k_{\perp}) \, 
{{d^2 k_{\perp}}\over{16 \pi^3}}
\end{equation}  
 is the  two-body contribution into the nonforward  parton density
 ${\cal F} (x,t)$.
Assuming  a Gaussian dependence on the transverse momentum $k_{\perp}$
(cf. \cite{bhl})    
\begin{equation}\Psi (x,k_{\perp}) =  \Phi(x) 
 e^{-k^2_{\perp}/2x \bar x \lambda^2} \,  , \label{11} 
 \end{equation}
we get 
\begin{equation}
{\cal F}^{(tb)} (x,t) = f^{(tb)}(x) e^{\bar x t /4 x \lambda^2 } \, , \label{8}
\end{equation}
where 
\begin{equation}
f^{(tb)}(x) = 
\frac{x \bar x  \lambda^2}{16 \pi^2} \, \Phi^2(x) 
= {\cal F}^{(tb)} (x,t=0)
\end{equation}
is the two-body part of the relevant parton density.
Within the light-cone approach, to get the total
result for either  usual $f(x)$
or 
nonforward parton densities   ${\cal F}(x,t)$,
one should 
add the contributions due to  higher Fock components.
By no means these contributions
are small, e.g., the  valence $\bar d u$  contribution
into the normalization of the $\pi^+$  form factor 
at $t=0$ is less than 25\% \cite{bhl}. 
In the absence of a formalism providing  explicit expressions
for an infinite tower of light-cone wave functions 
 we choose to treat  Eq.(\ref{8}) as a guide
for  fixing interplay between the $t$ and $x$ dependences
of NDs and propose to 
 model them by 
\begin{equation}
{\cal F}^a(x,t) = f_a(x) e^{\bar x t /4 x \lambda^2 }  = 
{{f_a(x)}\over{\pi x \bar x  \lambda^2}}\,
\int  \, e^{-(k^2_{\perp}+ (k_{\perp}+
\bar x r_{\perp})^2)/2x \bar x \lambda^2}
d^2 k_{\perp} \,  . \label{13}
\end{equation}
The functions  $f_a(x)$  here are the 
usual parton densities assumed to be 
taken from existing  parametrizations like GRV, MRS, CTEQ, etc.
In the $t=0$
limit (recall that $t$
is negative)  this model, by construction,  
satisfies the  first of  reduction formulas (\ref{3b}). 
 Within the Gaussian ansatz (\ref{13}), 
 the basic scale $\lambda$ specifies  the 
  average transverse momentum carried by the quarks.
   In particular, for valence quarks 
  \begin{equation} \langle  k^2_{\perp} \rangle ^a = 
  \frac{\lambda^2}{N_a}\int_0^1 
  x \bar x f_a^{val}(x)  \,  dx   \, , 
  \end{equation}
  where $N_u=2, N_d=1$ are the numbers of 
  the valence $a$-quarks in the
  proton.

  To fix the magnitude of  $\lambda$, we  
  use the second reduction formula in 
  (\ref{3b}) relating ${\cal F}^a(x,t)$'s
  to the $F_1(t)$ form factor.
  To this end, we take the following simple expressions for
  the valence distributions
  \begin{equation} f_u^{val} (x) = 1.89 \,
   x^{-0.4} (1-x)^{3.5} (1+6x) \, , \end{equation}
  \begin{equation} f_d^{val} (x) = 0.54 \, 
   x^{-0.6} (1-x)^{4.2} (1+8x) \, . \end{equation}
   \begin{figure}[htb]
\mbox{
   \epsfxsize=8cm
 \epsfysize=8cm
 \hspace{2cm}  
  \epsffile{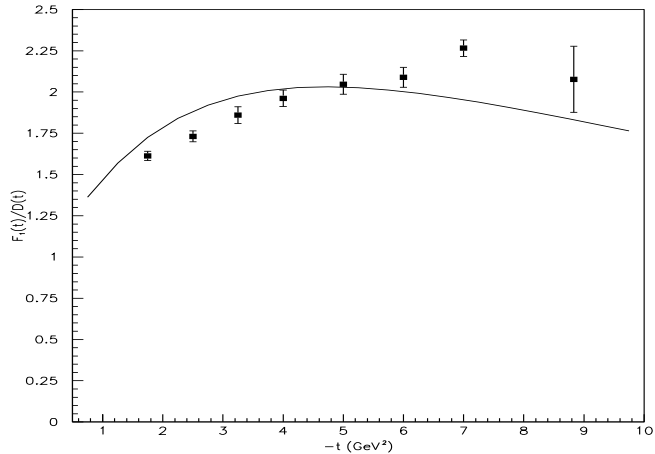}  }
{\caption{\label{fig:ff} Ratio $F_1^p(t)/D(t)$ 
of the $F_1^p(t)$ form factor to
the dipole fit $D(t) =1/(1-t/0.71\,{\rm GeV^2})^2$. Curve
is based on  Eq. (47) with $\lambda ^2 = 0.7 \,
{\rm GeV}^2$. Experimental
data are taken from ref.[34].
   }} 
\end{figure} 
\noindent They  closely reproduce the  relevant 
 curves given by the GRV parametrization \cite{grv} 
 at a low normalization point $Q^2 \sim 1$  GeV$^2$. 
 The best agreement between our  model 
 \begin{equation}
 F_1^{\rm soft}(t) =  \int_0^1 \left [ e_u\, f_u^{val} (x) +e_d
  \, f_d^{val} (x) \right ] e^{\bar x t / 4x \lambda^2} dx \label{14}
  \end{equation}
  and experimental data \cite{ff} in the
  moderately large $t$ region
   {1  GeV$^2$ $< |t|< 10$ GeV$^2$} is reached for 
   $\lambda^2 =0.7 \,$ GeV$^2$ (see Fig.\ref{fig:ff}). 
   This value gives a reasonable magnitude 
   \begin{equation}  \langle  k^2_{\perp} \rangle^u = (290 \,  {\rm MeV})^2 
   \  \   \  ,  \ \ \   \langle  k^2_{\perp} \rangle^d = (250 \, {\rm MeV})^2
   \end{equation}
   for the average transverse momentum of the valence $u$ and $d$ quarks
   in the proton.

 Similarly, building a model  for the 
parton helicity sensitive NDs ${\cal G}^a(x,t) $
one can take their  $t=0$ shape  from existing 
parametrizations for spin-dependent 
parton distributions $\Delta f_a(x)$
and  then fix the relevant $\lambda$ parameter by fitting
the $G_A(t)$ form factor. 
The case of hadron spin-flip distributions ${\cal K}^a(x,t)$ 
and ${\cal P}^a(x;t)$ is more complicated  
since the distributions $k_a(x)$, $p_a(x)$ are unknown.

At $t=0$, our model by construction gives a  correct 
normalization $F_1^p(t=0)=1$ for the form factor.
However, if one would try to find the derivative
$(d/dt)F_1^p(t)$ at $t=0$ by expanding the exponential 
$\exp [\bar x t/x \lambda^2]$ into the Taylor series under 
the integral (\ref{14}), one would get a divergent 
expression.  An analogous  problem is well known 
in applications of QCD sum rules to form factors 
at small $t$ \cite{iosmil,balyung,nestradsmall,belkogan}.
The divergence is related to the  
long-distance propagation of massless quarks in the $t$-channel.
Formally, this is revealed  by singularities  starting at $t=0$. 
However, $F_1^p(t)$ should not have singularities
for timelike $t$ up to $4 m_{\pi}^2$, with the $\rho$-meson 
peak at $t = m_{\rho}^2 \sim 0.6 \,{\rm GeV}^2$
being the most prominent feature of the $t$-channel spectrum.
Technically, the singularities of the original
expression are singled out into the  bilocal correlators
\cite{bal} which are substituted by  their 
realistic version with  correct spectral properties
(usually the simplest model with $\rho$ and $\rho '$
terms is used).  An important point
is that  such a modification is needed only 
when one calculates form factors in the  small-$t$ region:
for  $-t> 1\,{\rm GeV}^2$,  the  correction terms
should vanish  faster than any power of $1/t$ \cite{nestradsmall}. 
In our case, the maximum deviation of the curve for $F_1^p(t)$ given     
by  Eq.(\ref{14}) from the experimental data  in the small-$t$ region  
$-t< 1\, {\rm GeV}^2$  is 15$\%$. 
Hence, if one is willing  to tolerate such an inaccuracy,
one can use our model starting with $t=0$.  

Our curve  is within 5\% from the data  points \cite{ff} for
$1 \, {\rm GeV}^2 < -t < 6$ \, GeV$^2$ and does not deviate 
from them by more than 10\% up to 9 GeV$^2$.
Modeling the $t$-dependence by a more complicated formula
(e.g., assuming  a slower decrease at large $t$, and/or
choosing different $\lambda$'s for
$u$ and $d$ quarks and/or splitting NDs
into several components with different 
$\lambda$'s,  etc., see Ref.\cite{diehl} 
for an example of such an attempt) or changing the shape 
of parton densities $f_a(x)$ 
one can improve the quality of the fit
and extend  agreement with the data to higher $t$. 
Such a fine-tuning  is not our goal here. 
We just want to emphasize that 
a reasonable  description of the  $F_1(t)$ data in a wide region 
\mbox{1  GeV$^2$ $< |t|< 10$ GeV$^2$ } 
was obtained by fixing just a single parameter
 $\lambda$ reflecting the proton size. Moreover, we could 
fix $\lambda$ from the requirement that
$\langle  k^2_{\perp} \rangle \sim (300 \,  {\rm MeV})^2$
and present   our curve for $F_1(t)$ as a successful prediction 
of  the  model.    
We interpret this success as  an 
 evidence that the model
correctly catches the gross features of the 
underlying physics.

Since 
our   model implies  a Gaussian  
dependence on the transverse momentum,
it includes  only what is usually referred to as   
an overlap of soft wave functions.
It completely  neglects effects due to 
hard pQCD gluon exchanges  generating 
the power-law  $O( (\alpha_s / \pi)^2 /t^2)$
tail of the nonforward densities at large $t$.
It is worth pointing out here that though we take    nonforward 
densities  ${\cal F}^a(x,t)$ with  an exponential 
dependence on $t$, 
the $F_1(t)$ form factor in our model has 
a power-law asymptotics    $F_1^{\rm soft} (t) \sim (-4 
\lambda^2/t)^{n+1}$ 
dictated  by the   $(1-x)^n$ behavior of 
the parton densities for $x$ close to 1.
This connection arises because  
the integral (\ref{14}) over $x$ 
is dominated at large $t$ by  the region 
$\bar x \sim  4 \lambda^2 /|t|$. 
In other words,  the large-$t$ behavior of $F_1 (t) $ in our model
is governed by the Feynman mechanism \cite{feynman}. 
One should realize, however, that  the relevant scale
$4 \lambda^2 =2.8$ GeV$^2$ is rather large.
For this reason, when 
 $|t| < 10$ GeV$^2$,  
 it is premature to rely on asymptotic estimates 
 for the soft contribution. 
  Indeed, with $n=3.5$, the asymptotic estimate is  
 $F_1^{\rm soft} (t) \sim t^{-4.5}$,
in an apparent contradiction with 
the ability of our curve to follow 
the  dipole behavior.
The resolution of this paradox is very simple:
the maxima of nonforward densities  ${\cal F}^a(x,t)$ 
for $|t| < 10$ GeV$^2$ are 
at rather low $x$-values $x < 0.5$.
Hence, the $x$-integrals producing $F_1^{\rm soft} (t)$
are not dominated  by the $x \sim 1$ region yet and the 
asymptotic  estimates are not applicable:
the functional dependence of $F_1^{\rm soft} (t) $ in 
our model  is much more complicated than 
a simple power of $1/t$.    
 
  The fact that our    model  closely reproduces 
the experimentally 
observed 
dipole-like behavior of the proton form factor is a 
clear demonstration 
  that 
such a  behavior may have nothing to do with   the
quark counting rules $F_1^p(t) \sim 1/t^2$ \cite{brofar,mmt}
valid  for the
asymptotic behavior of 
the  hard gluon exchange 
contributions.
 Our explanation of the observed 
 magnitude and the $t$-dependence
 of  $F_1  (t)$ by a purely soft 
 contribution  is in  strong contrast   
 with that of the hard pQCD  approach to this problem.

\section{Wide-angle  Compton scattering}

With both photons real, it is not sufficient to have
large photon energy to ensure short-distance dominance:
large-$s$, small-$t$ region is strongly affected by
Regge contributions.
Hence, having large $|t| > 1 \, $GeV$^2$ is a  necessary
condition for revealing   short-distance  dynamics.

\begin{figure}[htb]
\mbox{
   \epsfxsize=12cm
 \epsfysize=3cm
 %\hspace{0.5cm}  
  \epsffile{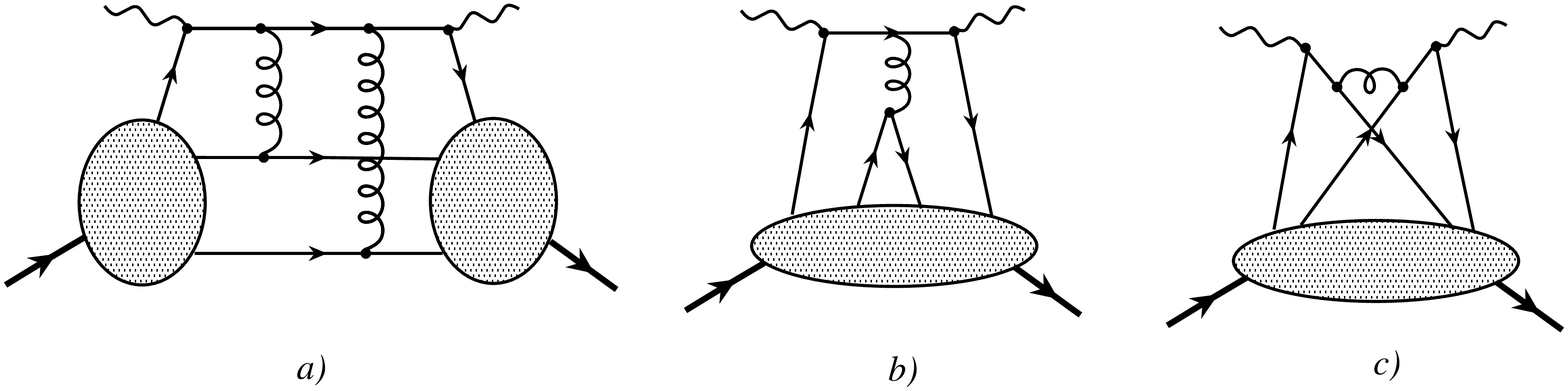}   }   
  \vspace{0.5cm}
{\caption{\label{fig:cosubno} Configurations 
involving double and single gluon exchange.
   }}
\end{figure}

The simplest contributions for the WACS amplitude 
are given by the $s$- and $u$-channel
handbag diagrams Fig.\ref{fig:cohan}b,c. 
The nonperturbative part in this case is given by the proton 
 DDs  
which determine the $t$-dependence of the total contribution.
Just like in the form factor case, 
the contribution dominating in the 
formal asymptotic limit $s,|t|, |u| \to \infty$,
is given
by diagrams corresponding to the pure SD regime,
see Fig.\ref{fig:cosubno}a. The hard subraph then 
involves two hard gluon exchanges which results in a  
suppression  factor $(\alpha_s/ \pi)^2 \sim 1/100$  
absent in  the handbag term. 
The total contribution  of all two-gluon exchnange  
contributions  was calculated by Farrar 
and Zhang
\cite{far} and then recalculated by  
 Kronfeld and Ni\v{z}i\'{c} \cite{kroniz}. 
 A sufficiently large 
 contribution is  only obtained if one uses  humpy DAs
 and   $1/k^2$ propagators with no finite-size effects included.
 Even with such propagators, 
 the WACS  amplitude 
calculated with the asymptotic DA is 
negligibly small \cite{vander} compared to existing  data. 
As argued in ref.\cite{realco}, 
 the    enhancements generated by 
the humpy DAs should not be taken at face  value both  
for form factors and  
wide-angle Compton scattering amplitudes. 
For these reasons, we ignore  hard contributions 
to the WACS  amplitude as negligibly small.

Another type of configurations containing  
 hard gluon exchange is  shown in Fig.\ref{fig:cosubno}b. There are  
also the diagrams with 
 photons coupled to different 
quarks (``cat's ears'', Fig.\ref{fig:cosubno}c). 
Such contributions have   both higher order 
and   higher twist.
This brings in the 
$\alpha_s/\pi $ factor  and  
 an  extra  $1/s$ suppression.
The latter is  partially compensated by  a slower fall-off 
of the four-quark DDs with $t$ since only one valence quark
should change its momentum.

\begin{figure}[htb]
\hspace{5mm} \mbox{
   \epsfxsize=5cm
 \epsfysize=4cm
  \epsffile{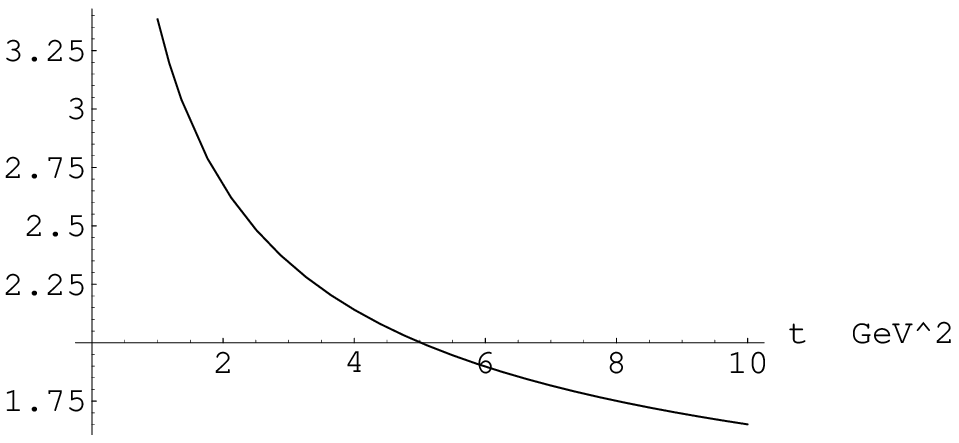} \hspace{1cm}    \epsfxsize=5cm
 \epsfysize=4cm
  \epsffile{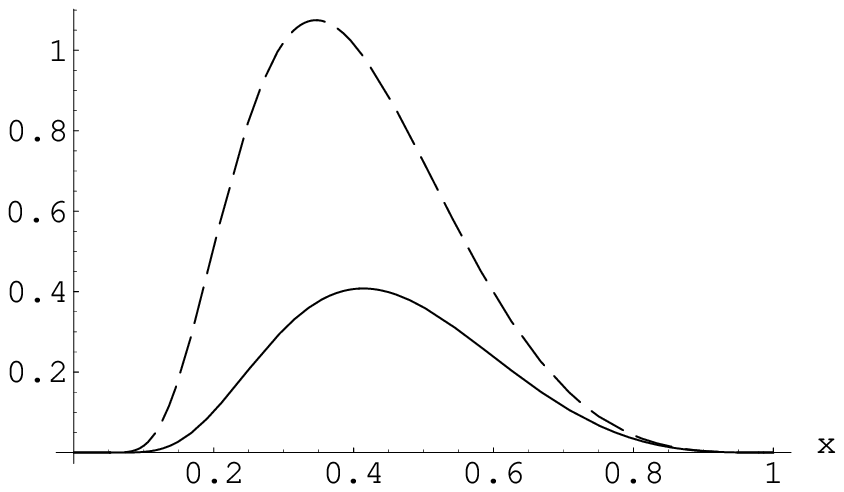} }
  \vspace{0.5cm}
{\caption{\label{fig:rcuux} $a)$ Ratio
$R_1^u(t)/F_1^u(t)$; $b)$ Functions  
${\cal F}^u (x;t)$ (solid line)
 and ${\cal F}^u (x;t)/x$ (dashed line) at  $t = - 2.5$ GeV$^2$. 
  }}
\end{figure}

For simplicity, we neglect all the suppressed terms 
and deal only with the handbag contributions Fig.\ref{fig:cohan}b,c 
in which 
the highly virtual quark propagator connecting the photon vertices 
  is convoluted with  proton DDs
  parametrizing the 
overlap of soft wave functions.  
Since  the  basic scale $4 \lambda^2$ 
characterizing the $t$-dependence of 
DDs in our model  is 2.8 GeV$^2$, while existing data    are 
all  at  momentum transfers $t$ below   5  GeV$^2$,
we deal with the region where the asymptotic estimate 
(Feynman mechanism) for the overlap
contribution is  not working yet.
In the coordinate representation, the sum of two 
handbag contributions to 
the Compton amplitude  can be written as 
\begin{eqnarray}
& & M^{\mu \nu} (p,p';q,q') = \sum_a e_a^2 \int  e^{-i(Qz)} 
\langle p'|
 ( \bar \psi_a (z/2) \gamma^{\mu} S^c(z) \gamma^{\nu} \psi_a (-z/2)
\nonumber \\ & & 
+ \bar \psi_a (-z/2) \gamma^{\nu} S^c(-z) 
\gamma^{\mu} \psi_a (z/2)) |p  \rangle \, 
d^4z
\end{eqnarray}
where $Q = (q+q')/2$ and
$S^c(z) = i \hat z/2  \pi^2 (z^2)^2 $ 
is the hard quark propagator 
(throughout, we use the 
``hat'' notation $\hat z \equiv z_{\alpha}\gamma^{\alpha}$).
The summation over the twist-0 longitudinal gluons
adds the usual gauge link between the $\bar \psi$,$ \psi$ fields
which we do not write down explicitly 
(gauge link disappears, e.g.,  in  the Fock-Schwinger gauge 
$z^{\alpha} A_{\alpha} (z) =0$).
Because of the symmetry of the problem, it is convenient to use 
$P = (p+p')/2$ (cf. \cite{ji}) and $r = p-p'$ as the basic 
momenta. 
Applying the  Fiertz transformation and  introducing the double  
distributions by 
\begin{eqnarray}
& & \langle p' |  \bar \psi_a (-z/2) \hat z
\psi_a (z/2) |p \rangle \, = 
\bar u(p') \hat z u(p) \, \int_0^1 \, dx \int_{-\bar x}^{\bar x} 
 \left [ e^{-i(kz)} { f}^a(x,\alpha ; t) \right. \nonumber \\
 & & \left. 
- e^{i(kz)} { f}^{\bar a} (x,\alpha  ;t) \right ] 
\, d\tilde y 
+ \frac1{4m_p}  \, \bar u(p') (\hat z \hat r - \hat r \hat z)  u(p)
 \label{17a} \\ & & \times 
\int_0^1 \, dx \int_{-\bar x}^{\bar x} \, 
\left [e^{-i(kz)} {k}^a(x,\alpha ;t) 
- e^{i(kz)} {k}^{\bar a} (x,\alpha ;t) \right ] 
\, d \alpha  + O(z^2) \  {\rm terms}\nonumber
 \end{eqnarray}
(we use here   the  shorthand notation 
$k  \equiv xP+ \alpha r/2$) and similarly for 
the  parton helicity sensitive operators
\begin{eqnarray} 
& & \langle  p'    |   \bar \psi_a(-z/2) \hat z \gamma_5
 \psi_a(z/2) 
 | p    \rangle   
=    \bar u(p')  \hat z  \gamma_5
 u(p) \,  \int_0^1 \, dx \int_{-\bar x}^{\bar x} 
 \left [  e^{-i(kz)}
 {g}^a(x,\alpha ;t) \right. \nonumber \\ 
& & \left.   +  e^{i(kz)}
{g}^{\bar a}(x, \alpha ;t)  \right ] 
 \,  d\alpha  
+
\, \frac{(rz)}{m_p} \,  \bar u(p')  \gamma_5
 u(p) \label{18}\\ & & \times    \int_0^1 \, dx \int_{-\bar x}^{\bar x} 
  \left [ e^{-i(kz)}
 {p}^a(x,\tilde y;t)  +  e^{i(kz)}
{ p}^{\bar a}(x,\alpha ;t)  \right ] 
 \,  d\alpha  + O(z^2) \  {\rm terms} \,  ,  \nonumber 
 \end{eqnarray} 
we arrive at  a leading-twist QCD parton picture 
with $\alpha$-DDs  serving as functions 
describing long-distance dynamics. 
The $\alpha$-DDs ${f}^a(x, \alpha ;t)$,  
 etc., are related to the original 
 $y$-DDs ${ F}^a(x,y;t)$  by the shift $y= (1-x+\alpha)/2$.
 Integrating  ${f}(x, \alpha ;t)$ over  
 $\alpha$ one obtains the same  
 nonforward densities  ${\cal F}(x ;t)$.
  The hard quark propagators  
for the $s$ and $u$ channel handbag diagrams 
in this picture  look like
 \begin{equation}
\frac{x \hat P + \alpha \hat r/2 + \hat Q}{(xP+\alpha r/2 +Q)^2} = 
\frac{ x \hat P + \alpha \hat r/2 + \hat Q}{x \tilde s -
(\bar x ^2  -\alpha^2)t/4 +x^2 m_p^2} 
\end{equation}
and
\begin{equation}
\frac{x \hat P + \alpha \hat r/2  - \hat Q}{(xP+\alpha r/2-Q)^2} =
\frac{x  \hat P+  \alpha \hat r/2 -\hat Q}{x \tilde u -
(\bar x ^2  -\alpha^2)t/4 +x^2 m_p^2} \, , 
\end{equation}
respectively. We denote $\tilde s =2(pq)=s-m^2$ and  
 $\tilde u =-2(pq')=u-m^2$.
 Since $\alpha$-DDs are even functions of $\alpha$
 \cite{lech}, the $\alpha \hat r$ terms 
 in the numerators can be dropped. 
It is legitimate to
keep $O(m_p^2)$ and $O(t)$ terms in the denominators: 
 the dependence of  hard propagators 
on target parameters $m_p^2$ and $t$ can be 
calculated exactly because of  the effect analogous to the 
$\xi$-scaling 
in DIS \cite{geopol} (see also \cite{rrnp}).
Note that the $t$-correction to hard propagators disappears 
in the large-$t$  limit dominated by the $x \sim 1$ integration.
The $t$-corrections are the largest for $y=0$. 
At this value and   for $x=1/2$  and $t=u$ (cm angle of 90$^{\circ}$), 
the $t$-term in the denominator of the most important
second propagator is 
only 1/8 of the $u$ term. This ratio increases to 1/3 for 
$x=1/3$. However, at nonzero $\alpha$-values, the $t$-corrections 
are smaller. Hence, the $t$-corrections in the denominators
of hard propagators can produce $10\% -20 \%$ effects  
and should be included in a complete analysis.     
Here, we consider an approximation in which 
 these terms are neglected 
and  hard propagators are given 
by $\tilde y$-independent 
expressions $(x \hat P + \hat Q)/x \tilde s$ 
and $(x \hat P + \hat Q)/x \tilde u$.
As a result, the  $\alpha$-integration acts only on the
  DDs ${f }(x,\alpha ;t)$ and 
converts them into  nonforward densities ${\cal F}(x,t)$. 
The latter    
would  appear then  through two types of integrals
 \begin{equation}
\int_0^1 
{\cal F}^a(x,t) \, {dx} \equiv F_1^a(t) \ \ {\rm and} \ \ \int_0^1 
{\cal F}^a(x,t) \, \frac{dx}{x} \equiv   R_1^a(t),  
\end{equation}
 and similarly for ${\cal K,G,P}$.
 The functions $F_1^a(t)$ are the flavor components 
 of the usual $F_1(t)$ form factor while $R_1^a(t)$
 are  the flavor components of a new form factor
  specific to the wide-angle Compton scattering.
  In the formal asymptotic limit $|t| \to \infty$, the $x$ integrals 
  for $F_1^a(t)$ and $R_1^a(t)$ 
  are both dominated in our model 
  by the $x \sim 1$ region: the large-$t$ 
  behavior  of these functions is governed 
  by the Feynman mechanism and their ratio tends to 1 as
  $|t|$  increases (see Fig.\ref{fig:rcuux}a). However, due to 
  large value of the effective scale $4 \lambda^2 =2.8$ GeV$^2$, 
  the  accessible momentum transfers
  $t < 5$  GeV$^2$ are very far from being asymptotic.

In Fig.\ref{fig:rcuux}b we plot  
   ${\cal F}^u (x;t)$ and ${\cal F}^u (x;t)/x$ at  $t = - 2.5$ GeV$^2$. 
  It is clear that the relevant integrals  are dominated
  by rather small $x$ values $x < 0.4$ 
  which results in a strong
  enhancement of $R_1^u(t)$ 
   compared to $F_1^u(t)$ for $|t| < 5$  GeV$^2$.
   Note also that the 
   $\langle p' | \ldots x \hat P \ldots |p \rangle $ 
   matrix elements  can produce only  $t$ as a large variable 
   while $\langle p' | \ldots  \hat Q \ldots |p \rangle $ 
   gives $s$. As a result, the enhanced form factors 
   $R_1^a(t)$ are accompanied by extra $s/t$ enhancement factors
   compared to the $F_1^a(t)$ terms. In the cross section,
   these enhancements are squared, i.e.,  
   the contributions due to the non-enhanced form factors   $F_1^a(t)$ 
  are always accompanied by $t^2/s^2$ factors
  which are smaller than 1/4 
  for cm angles 
  below 90$^{\circ}$. Because of double suppression,
    we neglect $F_1^a(t)$ 
  terms  in the present simplified 
  approach.

 The  contribution due to the ${\cal K}$ functions  
  appears through  the flavor components $F_2^a(t)$ of the $F_2(t)$
  form factor and their enhanced analogues $R_2^a(t)$. 
  The major part of contributions due to  the ${\cal K}$-type NDs 
  appears in the  combination 
  \mbox{$R_1^2(t)-(t/4m_p^2)R_2^2(t)$.}
  Experimentally, $F_2(t)/F_1(t)\approx 1 \,{\rm GeV}^2/|t|$.
  Since $R_2/F_2 \sim R_1/F_1 \sim 1/ \langle x  \rangle $, 
  $R_2(t)$ is similarly suppressed compared 
  to $R_1(t)$,  and we  neglect contributions 
  due to the $R_2^a(t)$ form factors.  
  We also neglect here the terms with 
  another spin-flip distribution  ${\cal P}$  related
  to the pseudoscalar form factor $G_P(t)$ which is dominated 
  by the $t$-channel pion exchange.  Our  calculations
  show that the contribution due to  
  the parton helicity sensitive densities  ${\cal G}^a$ 
  is suppressed by the factor $t^2/2s^2$ compared to that due to the 
  ${\cal F}^a$ densities. This factor only reaches
  1/8  for the cm angle of  
  90$^{\circ}$,  and hence the ${\cal G}^a$ contributions are not 
  very significant  numerically. For simplicity, we 
  approximate  ${\cal G}^a(x,t) $ by ${\cal F}^a(x,t)$. 
After these approximations,
the WACS  cross section is given by the product  
 \begin{equation}
 \frac{ d \sigma}{dt}  \approx \frac{2 \pi \alpha^2}{\tilde s^2} 
 \left [  \frac{(pq)}{(pq')} +  \frac{(pq')}{(pq)}
  \right ] \,  R_1^2(t) \,  , 
 \end{equation}
 of  the  Klein-Nishina  cross section 
(in which we dropped  $O(m^2)$ and $O(m^4)$ terms) and 
 the square of the $R_1(t)$ form factor 
  \begin{equation} 
 R_1(t) =  \sum_a e_a^2  \left [R_1^a (t) + R_1^{\bar a} (t) \right ] \, .
 \end{equation}
 In our model, $R_1(t)$ is given by 
 \begin{equation} 
R_1(t) =   \int_0^1 
\biggl [ e_u^2 \, f_u^{val} (x) +
  e_d^2  \, f_d^{val} (x) + 2( e_u^2+e_d^2+e_s^2) 
  \, f^{sea} (x) \biggr ] 
  e^{\bar x t / 4x \lambda^2} \frac{dx}{x} \, . 
\end{equation}
 We included here  the 
 sea   distributions assuming that they are all
  equal $f^{sea} (x)
=f^{sea}_{u,d,s} (x)= f_{\bar u, \bar d,\bar s} (x)  $ 
and using  a  simplified parametrization 
\begin{equation}
f^{sea} (x) = 0.5 \,  x^{-0.75} (1-x)^7
\end{equation}
which  accurately reproduces  
 the  GRV formula for $Q^2 \sim 1$ GeV$^2$.
 Due to  suppression  of the small-$x$ region
 by the exponential $\exp [\bar x t / 4x \lambda^2]$,
 the sea quark contribution is rather 
 small ($\sim 10 \%$) even for $-t \sim  1$   GeV$^2$ and 
 is invisible  for  $-t > 3 $   GeV$^2$.

\begin{figure}[htb]
\mbox{
   \epsfxsize=8cm
 \epsfysize=11cm
 \hspace{2cm}  
  \epsffile{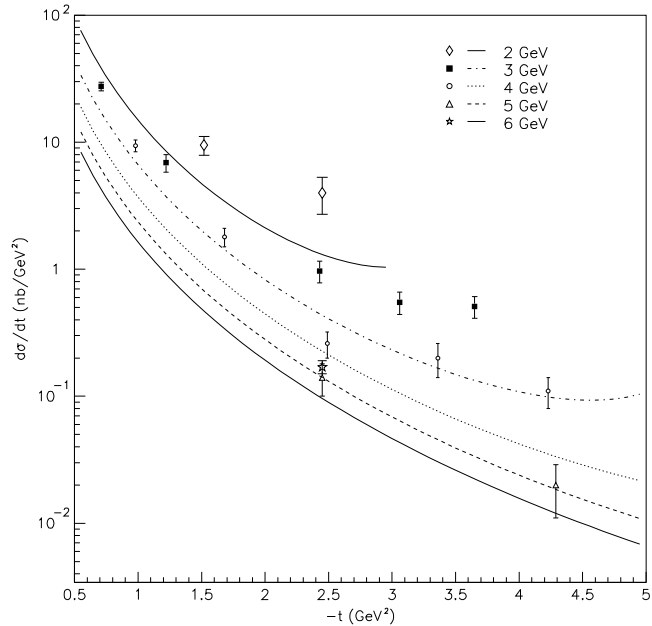}  }
  \vspace{-1.2cm}
{\caption{\label{fig:rct} WACS cross section versus $t$: 
comparison of results 
based on Eq.(55) with experimental data.
   }}
\end{figure}

Comparison with existing data \cite{schupe} 
is shown in Fig.\ref{fig:rct}.  Our curves  follow the data pattern
but are systematically lower  by a factor of  2,
with disagreement becoming more pronounced 
 as  the scattering angle  increases.
Since we neglected several terms each capable 
of producing up to a $20 \%$ correction in  the amplitude, we consider
the agreement between our curves and the data 
as encouraging. The most important corrections which should 
be included in a more detailed investigation
are the $t$-corrections in the denominators of 
hard propagators and contributions due to the ``non-leading''
${\cal K, G,P}$ nonforward densities.
The latter, as noted above, are usually accompanied 
by $t/s$ and $t/u$ factors, i.e., their contribution 
becomes  more significant at larger angles. 
The $t$-correction in the most important hard propagator term  
$1/[x \tilde u - (\bar x^2 - \alpha^2)t/4 +x^2 m_p^2]$
also enhances the amplitude at large angles.

\begin{figure}[htb]
\mbox{
   \epsfxsize=8cm
 \epsfysize=11cm
 \hspace{2cm}  
  \epsffile{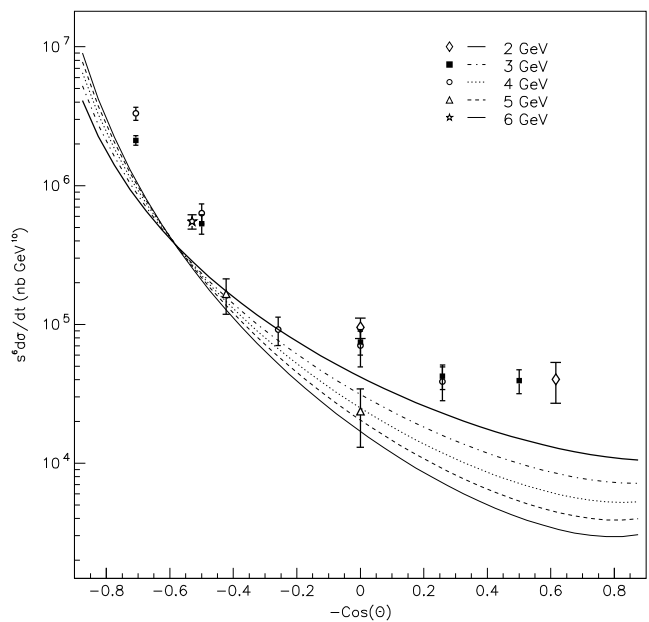}  }
  \vspace{-1.2cm}
{\caption{\label{fig:rctheta}  Angular dependence of 
the combination $s^6 (d \sigma /dt)$.
   }}
\end{figure}

The angular dependence of our results for the combination
$s^6 (d \sigma /dt)$ is shown on Fig.\ref{fig:rctheta}.
All the curves for initial photon ehergies 2,3,4,5 and 6 GeV
intersect each other at  $\theta_{\rm cm} \sim 60^{\circ}$.
This  is in good agreement with experimental data
of ref.\cite{schupe} where the differential 
cross section at fixed cm angles was fitted by powers of $s$:
$d \sigma /dt \sim s^{-n (\theta)}$ with 
$n^{\rm exp}(60^{\circ}) = 5.9 \pm 0.3$. 
Our curves correspond to $n^{\rm soft}(60^{\circ}) \approx 6.1$
and $n^{\rm soft}(90^{\circ}) \approx 6.7$ which also agrees 
with the experimental result $n^{\rm exp}(90^{\circ}) = 7.1 \pm 0.4$.

This can be compared with the scaling behavior
of the asymptotic  hard contribution: 
modulo logarithms contained in the $\alpha_s$ factors,
they have    a universal angle-independent power
$n^{\rm hard}  (\theta) =6$.
For $\theta_{\rm cm}  = 105^{\circ}$, the experimental result
based on just two  data points is $n^{\rm exp}(105^{\circ}) = 6.2 \pm 1.4$,
while our model gives $n^{\rm soft}(105^{\circ}) \approx 7.0$.
Clearly, better data are needed to draw any conclusions here.

\begin{figure}[htb]
\mbox{
   \epsfxsize=8cm
 \epsfysize=11cm
 \hspace{2cm}  
  \epsffile{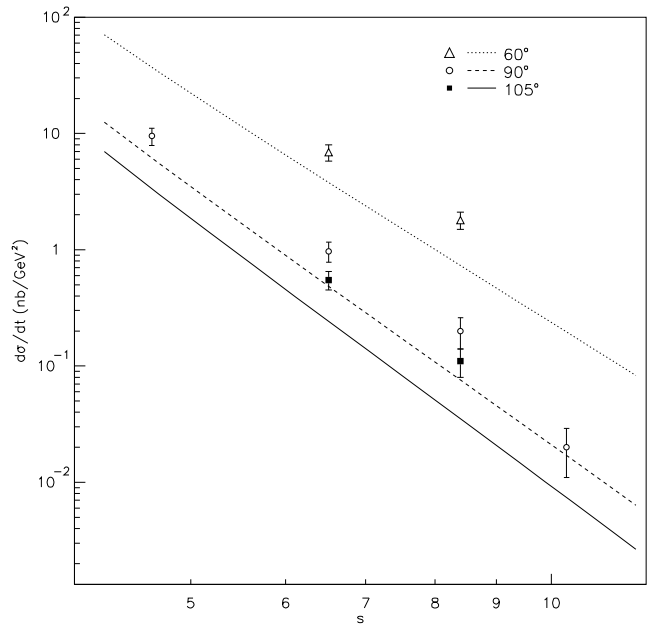}  }
  \vspace{-1.2cm}
{\caption{\label{fig:rcs} $s$-dependence of 
the combination  $s^6 d \sigma /dt $ 
 for $\theta = 60^{\circ}$
(dotted line),  
$\theta = 90^{\circ}$ (dashed line)
and $\theta = 105^{\circ}$ (solid line).
   }}
\end{figure}

\section{Conclusions}

The hard exclusive electroproduction processes 
provide   new  information about 
hadronic structure accumulated in 
skewed  parton distributions.  The SPDs are   
universal  hybrid functions having the 
properties of parton densities,  hadronic form factors 
and distribution
amplitudes.    They give a unified description of  
various hard  exclusive and inclusive reactions. 
The basic supplier of information
about skewed parton distributions is 
deeply virtual Compton scattering   which 
offers a remarkable 
example of Bjorken scaling 
phenomena  in 
exclusive processes.  
Furthermore, wide-angle real Compton scattering is an ideal tool
to test angle-dependent scaling laws characteristic 
for soft overlap mechanism.

\section*{Acknowledgements}

I am grateful to  A. Bialas and M. Praszalowicz  
for hospitality in  Zakopane and support. 
I thank  K. Golec-Biernat for
discussions.  This work was supported 
by the U.S. Department of Energy under Contract No.
DE-AC05-84ER40150.

\end{document}